\documentclass[twocolumn,pra,showpacs]{revtex4}
\usepackage{dcolumn,times}
\usepackage{epsfig,color,psfrag,amssymb,amsmath,amsthm,amsbsy}
\newcommand{\GPE}{Gross-Pitaevskii equation}
\newcommand{\BEC}{Bose-Einstein condensate}

\newcommand{\Label}[1]{\label{#1}}

\newlength{\hgt}
\newcommand{\FIG}[1]{CompressedFigs/#1}

\begin{document}

\title{Quantum turbulence and correlations in Bose-Einstein condensate
collisions}

\author{A. A. Norrie, R. J. Ballagh, and C. W. Gardiner}

\affiliation{Department of Physics, University of Otago, Dunedin, New Zealand}

\begin{abstract}
We investigate numerically simulated collisions between experimentally
realistic Bose-Einstein condensate wavepackets, within a regime where highly
populated scattering haloes are formed. The theoretical basis for this work is
the truncated Wigner method, for which we present a detailed derivation, paying
particular attention to its validity regime for colliding condensates. This
paper is an extension of our previous Letter \cite{Norrie2005a}, and we
investigate both single-trajectory solutions, which reveal the presence of
quantum turbulence in the scattering halo, and ensembles of trajectories, which
we use to calculate quantum-mechanical correlation functions of the field.
\end{abstract}


\maketitle


\section{Introduction}

In the same way as a classical electromagnetic field obeying Maxwell's
equations arises as an assembly of photons all in the same quantum state, a
\BEC, composed of Bosonic atoms all in the same quantum state, behaves very
much like a classical field $\Psi({\bf x},t)$, whose equation of motion is the
\GPE
\begin{eqnarray*}
\Label{GPE in coordinate space}
i\hbar \frac{\partial \Psi \left( \mathbf{x},t\right) }{\partial t}=
\left[ -\frac{\hbar ^{2}\nabla ^{2}}{2m}+\frac{4\pi \hbar ^{2}a}{m}
\left| \Psi \left( \mathbf{x},t\right) \right| ^{2}\right] 
\Psi \left( \mathbf{x},t\right).
\end{eqnarray*}
The \GPE\ has been extraordinarily successful in describing a wide
range of the phenomena associated with \BEC s---nevertheless, there
are phenomena in which the quantized nature of this field is
important. For example, when two Bose-Einstein condensates collide at
a sufficiently high velocity, a \emph{halo} of elastically scattered
atoms is produced \cite{Chikkatur2000a,Katz2002a,Vogels2002a}.  The
\GPE\ with initial conditions corresponding to two \BEC s does not
predict this scattering---it is a direct effect of the fact that the
quantized field consists of interacting particles, and at its most
elementary level, this halo is simply the result of elastic scattering
of the constituent particles in the \BEC.

A description of this phenomenon by means of  the phenomenological 
inclusion of loss terms \cite{Band2000a}, in a way reminiscent of the Boltzmann 
equation, was successful for collisions of smaller condensates, but 
seriously under-estimated the scattering for higher condensate 
densities. This underestimation arises because at high condensate
densities, the final states into which the scattering occurs can
become sufficiently highly occupied to cause \emph{Bosonic
stimulation}, which a simple Boltzmann treatment cannot produce.
On the other hand, a treatment in terms of linearized 
quantum field theory  \cite{Bach2002a,Yurovsky2002a} can deal with 
the Bosonic stimulation effects of highly occupied final states, 
but, being linearized, can treat neither the effects of large
depletion, nor the essentially classical nonlinearity effects so well 
handled by the \GPE.

An alternative method is the \emph{truncated Wigner method}.  This is
a treatment of quantum field theory in which quantum mechanics is
simulated by a classical random process.  The method is approximate,
but nevertheless very useful, both for quantum optical systems, where
the field under consideration is the electromagnetic field, and for
Bose-Einstein condensates, where the matter-wave field is under
consideration, and reproduces many quantum mechanical features, such
as quantum correlation functions, of these systems.
The application to \BEC s has been developed by several groups
\cite{Steel1998a,Stoof1999a,Duine2001a,Stoof2001a,Davis2001b,Davis2001a,
Sinatra2001a,Goral2001a,Gardiner2002a,Davis2002a,Goral2002a,Schmidt2003a}
with considerable success.  

In qualitative terms, the truncated Wigner method provides a
description of quantum field theory based on the \GPE in which quantum
mechanical vacuum fluctuations are simulated by by adding appropriate
classical fluctuations in addition to the coherent field of the
\emph{initial state} of the \BEC. These amount to half a quantum per
degree of freedom, corresponding to the zero point energy of the
harmonic oscillator which represents each mode of the field---the
precise way in which this is done is presented in
Sect.\ref{SSSection: initial states}. %
The elastic scattering effects which are not produced
directly by a solution of the \GPE\ for a coherent initial condensate field
now appear as four-wave mixing between the coherent condensate field
and the simulated vacuum fluctuations.

Since the number of modes in a \emph{local} quantum field theory is infinite,
the addition of a half a quantum per mode does in principle introduce
a infinite density of vacuum fluctuations.  Nevertheless, this does
not invalidate the method.  In practice what is used for the
description of a \BEC\ is an \emph{effective field
theory}\cite{Braaten1999a,Andersen2004a}, valid only on a rather
coarse spatial scale.  Because of this, one can use a quantum field
theory which contains only modes up to a certain cutoff value of the
momentum, with an appropriately renormalized interaction.  In a
practical implementation of a cutoff field theory,
one uses the standard contact interaction, with strength
proportional to the scattering length, with corrections which depend
on the size of the momentum cutoff.  The density of added vacuum
fluctuations is then quite finite, but cutoff dependent.  The cutoff
dependent effects of the vacuum fluctuations are then compensated by
the cutoff dependent interaction strength.  In simulations we must
introduce this cutoff explicitly---in practice, for the choices of
parameters we make, the cutoff corrections are 0.5\% ---see 
Sects.\ref{sssection: validity of the effective field theory},
\ref{ssection: Truncated Wigner method validity criteria}.

A related approach, the \emph{positive-P} method
\cite{Steel1998a,Drummond1999a}, (and its generalization, the
\emph{gauge positive-P} method) has also proven useful in describing
Bose-condensed systems.  Both the truncated Wigner and positive-P
methods are examples of the general class of ``classical field
methods'' well known in quantum optics \cite{Gardiner2000a}.  The
positive-P methods are in principle exact, but are still difficult to
implement for the kind of system we are considering here.  However,
technical progress is currently being made
\cite{Deuar2006a,Deuar2005a}, and there does seem to be 
real promise that these exact methods will become practical. 

We recently  applied the truncated Wigner method to the problem of
colliding condensates \cite{Norrie2005a}, showing that it was feasible
to simulate a realistic three-dimensional system and produce
quantitative results.
The aim of this paper is to expand upon and extend the treatment presented in
there. To this end we provide here: 

\begin{itemize}
 \item[a)] A detailed derivation of the truncated Wigner method for
 colliding condensates, including a heuristic demonstration of how we
 can justify the neglect of terms arising in a Wigner function
 treatment of the problem, setting up the foundation of the method.
 We show in Sects.\ref{validity}--\ref{critique} that for this kind of problem, the
 validity criterion for the method is that the density of the
 condensate (in co-ordinate space) must be very large compared with
 the density of the added quantum fluctuations.

\item[b)] A treatment of the computational aspects of the problem,
which are rather subtle.  The main feature to note is that the cutoff
necessary in the effective field theory cannot be simply provided by
the fineness of the spatial grid used for computations.  Rather,
in order to avoid aliasing, it must be provided explicitly by means of
a projector.

\item[c)] The evaluation of averages using full ensemble computations.
In \cite{Norrie2005a} we evaluated averages using single computational
runs, and averaging over regions of space where symmetry indicated the
physics was the same.  The results of our ensemble methods can provide
additional information on coherence properties of the final states.

\item[d)]  Results which can be compared with feasible experiments.
At present there are no experimental data which can be
quantitatively compared with the results of this paper. The work was
motivated by the observation of a strong halo in \cite{Vogels2002a},
but no detailed measurements were made on this halo which could be
compared with theory---furthermore, the parameter regime is not fully 
within the range of validity of our methodology.  
The work reported in \cite{Katz2005a} has adapted our theoretical
methodology to analyze a related experiment, and got good agreement.  
However, the calculation done was only two-dimensional, and therefore
can only be regarded as indicative.

For this paper we have chosen a parameter regime which is
experimentally attainable, which is in the region where strong Bosonic
stimulation is important, and which is fully within the region of
validity of our methodology.  In Sect.\ref{Section: system of
interest} we give the values of appropriate parameters for both sodium
and rubidium condensates.  The numerical results are presented in SI
units for a sodium condensate, and should be directly verifiable
experimentally.
\end{itemize}

\section{Truncated Wigner method}

In dilute Bose gases the appropriate Schr\"odinger picture Hamiltonian is
\begin{eqnarray}
\label{Full Hamiltonian 1}
\hat{H} \hspace{-0.1cm} & = & \hspace{-0.2cm}
\int d{\bf x} \, \hat{\Psi} ^{\dagger} \left( {\bf x} \right)
\left\{ - \frac{\hbar^2 \nabla^2}{2m} + U_{\rm ext} \left( {\bf x} \right)
\right. \nonumber \\
&& \hspace{-0.5cm} \left. \mbox{} + \frac{1}{2} \int d{\bf x}' \,
\hat{\Psi}^{\dagger} \left( {\bf x}' \right)
U_{\rm 2b} \left( {\bf x} - {\bf x}' \right)
\hat{\Psi} \left({\bf x}' \right) \right\} \hat{\Psi} \left({\bf x} \right).
\end{eqnarray}
Here the external potential is $U_{\rm ext} \left( {\bf x} \right)$ and
pairwise interactions between the bosons are characterized by the two-body
scattering potential $U_{\rm 2b} \left( {\bf x - x}' \right)$. The
second-quantized field operator $\hat{\Psi} \left( {\bf x} \right)$ annihilates
a particle from position ${\bf x}$ and obeys the equal time commutation
relations for identical bosons
\begin{eqnarray}
\label{Full field operator commutations}
\left[ \hat{\Psi} \left( {\bf x} \right),
\hat{\Psi} \left( {\bf x}' \right) \right] & = &
\left[ \hat{\Psi}^{\dagger} \left( {\bf x} \right),
\hat{\Psi}^{\dagger} \left( {\bf x}' \right) \right] = 0
\nonumber \\
\left[ \hat{\Psi} \left( {\bf x} \right),
\hat{\Psi}^{\dagger} \left( {\bf x}' \right) \right] & = &
\delta \left( {\bf x} - {\bf x}' \right),
\end{eqnarray}
where $ \delta \left( {\bf x} \right) $ is the three-dimensional Dirac delta
function.

\subsection{Effective field theory}
\label{ssection: effective field theory}

\subsubsection{Restricted basis field}

We now decompose the field operator onto a single-particle basis
\begin{equation}
\label{Full field operator}
\hat{\Psi} \left( {\bf x} \right) = \sum_{j} \psi_j \left( {\bf x}
\right) \hat{a}_j,
\end{equation}
where the mode operators obey the usual bosonic commutation relations
\begin{equation}
\label{Mode operator commutations}
\left[ \hat{a}_i, \hat{a}_j \right] =
\left[ \hat{a}_i^{\dagger}, \hat{a}_j^{\dagger} \right] = 0, \hspace{0.5cm}
\left[ \hat{a}_i, \hat{a}_j^{\dagger} \right] = \delta_{i,j}.
\end{equation}
By choosing the basis set to be the orthonormal eigenstates of the
non-interacting portion of the Hamiltonian, Eq.~(\ref{Full Hamiltonian 1}),
\emph{i.e.}
\begin{equation}
\left\{ -\frac{\hbar^2 \nabla^2}{2m} + U_{\rm ext}
\left( {\bf x} \right) \right\} \psi_j \left( {\bf x} \right)
= \hbar \omega_j \psi_j \left( {\bf x} \right),
\end{equation}
the full Hamiltonian, Eq.~(\ref{Full Hamiltonian 1}) can be rewritten as
\begin{equation}
\label{Full Hamiltonian 2}
\hat{H} = \sum_{j} \hbar \omega_j \hat{a}^{\dagger}_j \hat{a}_j
+ \frac{1}{2}\sum_{jrst} \left \langle jr \left| U_{\rm 2b} \right| st \right
\rangle \hat{a}_j^{\dagger} \hat{a}_r^{\dagger} \hat{a}_s \hat{a}_t.
\end{equation}

It is usual to simplify the Hamiltonian, Eq.~(\ref{Full Hamiltonian 2}), by
using an \emph{effective field theory}, obtained by eliminating higher energy
modes, whose time-dependence is so rapid as to be unobservable in experiments
on ultra cold gases. This kind of procedure has a long history, and takes many
different forms.  The description most appropriate to our methodology was given
by Morgan \cite{Morgan2000a}, and divides the modes $j$ into two sets, low-
($L$) and high-energy ($H$) subspaces depending on whether $\hbar\omega_j$ is
less than or greater than a certain \emph{boundary energy} $\varepsilon_{\rm
cut}$. Provided  $\varepsilon_{\rm cut}$ is sufficiently small, the effective
Hamiltonian describing the $M$ low-energy modes can be found from
Eq.~(\ref{Full Hamiltonian 2}) to be
\begin{equation}
\label{Effective Hamiltonian 1}
\hat{H}_{\rm eff} = \sum_{j \in L} \hbar \omega_j \hat{a}^{\dagger}_j \hat{a}_j
+ \frac{U_0}{2} \sum_{jrst \in L} \left \langle jr | st \right \rangle
\hat{a}_j^{\dagger} \hat{a}_r^{\dagger} \hat{a}_s \hat{a}_t,
\end{equation}
where the interaction parameter is defined as $U_0 \equiv 4\pi \hbar^2 a / m$
for the $s$-wave scattering length $a$.

\subsubsection{Validity of the effective field theory}
\label{sssection: validity of the effective field theory}

In order for $\hat{H}_{\rm eff}$ to accurately describe the low-energy
subspace, $\varepsilon_{\rm cut}$ must be chosen so that the evolution of the
high-energy modes is rapid compared to the evolution of the low-energy system,
and $L$ must include enough modes to adequately describe the system dynamics.
For the colliding systems we treat here, the low-energy subspace must be
sufficiently large to contain all the modes required to represent the two
colliding condensates and the scattering halo, as well as all the modes that
can be directly involved in scattering events with those (condensate and halo)
modes. In principle, this may conflict with the requirement that
$\varepsilon_{\rm cut}$ be small --- however in the cases we consider here, the
error is less than a few percent, as we describe in section 
\ref{Section: system of interest}.

\subsubsection{Projector representation}

It is useful to have some formal manner of decomposing arbitrary objects, such
as field operators or wavefunctions, into components on either side of the
boundary energy. To this end we define the orthogonal projection operators
(projectors)
\begin{equation}
\mathcal{P} \equiv \sum_{j \in L} \left| j \right \rangle \left \langle j
\right|, \hspace{0.5cm}
\mathcal{Q} \equiv \sum_{j \in H} \left| j \right \rangle \left \langle j
\right|,
\end{equation}
where $\mathcal{P}$ and $\mathcal{Q}$ act, respectively, as projectors onto the
low- ($L$) and high-energy ($H$) subspaces. As these two subspaces completely
span the infinite mode space, the projectors $\mathcal{P}$ and $\mathcal{Q}$
satisfy the closure relation $\mathcal{P} + \mathcal{Q} = 1$.

In its coordinate space form, the low-energy projector acts on the arbitrary
function $f \left( {\bf x} \right)$ as 
\begin{equation}
\label{Spatial P operator}
\mathcal{P} \left[ f \left( {\bf x} \right) \right] =
\sum_{j \in L} \psi_j \left( {\bf x} \right)
\int d{\bf x}' \psi_j^* \left( {\bf x}' \right) f \left( {\bf x}' \right).
\end{equation}
Applying this to the field operator given by Eq.~(\ref{Full field operator})
returns the \emph{restricted field operator}
\begin{equation}
\label{Restricted field operator}
\mathcal{P} \left[ \hat{\Psi} \left( {\bf x} \right) \right] =
\sum_{j \in L} \psi_j \left( {\bf x} \right) \hat{a}_j
\equiv \hat{\Psi}_{\mathcal{P}} \left( {\bf x} \right),
\end{equation}
which is the component of the total field operator acting within the low-energy
subspace. Because of the restricted nature of $\hat{\Psi}_{\mathcal{P}}$, the
commutation relations given by Eq.~(\ref{Full field operator commutations}) no
longer apply. Rather it can be shown, using the mode operator commutation
relations (\ref{Mode operator commutations}), that the restricted field
operator obeys the equal time relations
\begin{eqnarray}
\left[ \hat{\Psi}_{\mathcal{P}} \left( {\bf x} \right),
\hat{\Psi}_{\mathcal{P}} \left( {\bf x}' \right) \right] & = &
\left[ \hat{\Psi}_{\mathcal{P}}^{\dagger} \left( {\bf x} \right),
\hat{\Psi}_{\mathcal{P}}^{\dagger} \left( {\bf x}' \right) \right] = 0
\nonumber \\
\left[ \hat{\Psi}_{\mathcal{P}} \left( {\bf x} \right),
\hat{\Psi}_{\mathcal{P}}^{\dagger} \left( {\bf x}' \right) \right] & = &
\delta_{\mathcal{P}} \left( {\bf x},{\bf x}' \right), \hspace{1cm}
\end{eqnarray}
where we have defined the \emph{restricted delta function}
\begin{equation}
\label{Restricted delta function}
\delta_{\mathcal{P}} \left( {\bf x},{\bf x}' \right) \equiv
\sum_{j \in L} \psi_j^* \left( {\bf x}' \right)
\psi_j \left( {\bf x} \right).
\end{equation}
Unlike the true (Dirac) delta function, the restricted delta function is
spatially nonlocal, where the range of this nonlocality scales as
$\varepsilon_{\rm cut}^{-1/2}$.

\subsection{Wigner function evolution}

Let us define the density operator of the restricted basis field to be
$\rho \left( t \right)$, whose time evolution, using Eq.~(\ref{Effective
Hamiltonian 1}), is given by
\begin{eqnarray}
i\hbar \frac{d \rho}{dt} & = & \left[
\hat{H}_{\rm eff}, \rho \right] \\
\label{Von Neumann equation}
& = &
\sum_{j \in L} \hbar \omega_j \left( \hat{a}_j^{\dagger} \hat{a}_j \rho -
\rho \hat{a}_j^{\dagger} \hat{a}_j \right) \nonumber \\
&& \hspace{-1cm} \mbox{}
+ \frac{U_0}{2} \sum_{jrst \in L}
\left \langle jr| st \right \rangle \left( \hat{a}_j^{\dagger}
\hat{a}_r^{\dagger} \hat{a}_s \hat{a}_t \rho -
\rho \hat{a}_j^{\dagger} \hat{a}_r^{\dagger} \hat{a}_s \hat{a}_t \right),
\end{eqnarray}

The formulation of the truncated Wigner method is made using a multimode
\emph{Wigner function representation} of the density operator $\rho \left( t
\right)$. The full details of how this is used are given in
\cite{Gardiner2000a}, and in brief are as follows. For a single mode, the
Wigner function $W \left( \alpha, \alpha^*,t \right)$ is defined in terms of
the Wigner characteristic function
\begin{equation}
\label{quantum characterstic function}
\chi_W \left( \lambda,\lambda^* \right) \equiv {\rm Tr} \left\{ \rho
\exp \left( \lambda \hat{a}^{\dagger} - \lambda^* \hat{a} \right) \right\},
\end{equation}
as a Fourier transform
\begin{equation}
\label{wig1}
W(\alpha,\alpha^*) = \frac{1}{\pi^2} \int d^2 \lambda \,
\exp \left( -\lambda \alpha^* + \lambda^* \alpha \right)
\chi_W(\lambda,\lambda^*).
\end{equation}
The Wigner function exists for any density operator, and its moments are equal 
to those of the symmetrized operator products:
\begin{eqnarray}
\label{wig2}
\left \langle \left\{ \hat{a}^r \left( \hat{a}^{\dagger} \right)^s
\right\}_{\rm sym} \right \rangle & = &
\int d^2 \alpha \, \alpha^r \left( \alpha^* \right)^s
W \left( \alpha,\alpha^* \right) \\
& \equiv & \left \langle \alpha^r \left( \alpha^* \right)^s \right \rangle_W.
\end{eqnarray}
The Wigner function is not guaranteed to be positive, but often is, and in 
these cases it behaves like a probability distribution for the variables
$\alpha$, $ \alpha^*$.

The action of the mode operators $\hat{a}, \hat{a}^{\dagger}$ on
the density operator can be expressed as the action of differential operators
on the Wigner function using the \emph{operator correspondences}, which follow
from Eq.~(\ref{wig1})

\begin{subequations}
\begin{eqnarray}
\hat{a} \rho \left( t \right) & \leftrightarrow &
\left( \alpha + \frac{1}{2} \frac{\partial}{\partial \alpha^*} \right)
W \left( \alpha,\alpha^*,t \right) \\
\hat{a}^{\dagger} \rho \left( t \right) & \leftrightarrow &
\left( \alpha^* -\frac{1}{2} \frac{\partial}{\partial \alpha} \right)
W \left( \alpha,\alpha^*,t \right) \\
\rho \left( t \right) \hat{a} & \leftrightarrow &
\left( \alpha - \frac{1}{2} \frac{\partial}{\partial \alpha^*} \right)
W \left( \alpha,\alpha^*,t \right) \\
\rho \left( t \right) \hat{a}^{\dagger} & \leftrightarrow &
\left( \alpha^* +\frac{1}{2} \frac{\partial}{\partial \alpha} \right)
W \left( \alpha,\alpha^*,t \right).
\end{eqnarray}
\end{subequations}

The extension to many modes is straightforward.

\begin{widetext}
For the restricted basis von Neumann equation, Eq.~(\ref{Von Neumann
equation}), we find using these operator correspondences that the evolution of
the multimode Wigner function $W \left( \left\{ \alpha_j,\alpha_j^* \right\},t
\right)$ is given by
\begin{eqnarray}
\label{Wigner function evolution full}
i \hbar \frac{\partial W}{\partial t} & = & - \sum_{j \in L}
\hbar \omega_j \left( \frac{\partial}{\partial \alpha_j} \alpha_j
- \frac{\partial}{\partial \alpha_j^*} \alpha_j^* \right) W \nonumber \\
&& \mbox{} - U_0 \sum_{jrst \in L} \left \langle jr | st \right \rangle
\left( \frac{\partial}{\partial \alpha_j} \alpha_t -
\frac{\partial}{\partial \alpha_t^*} \alpha_j^* \right)
\left( \alpha_r^* \alpha_s - \delta_{r,s} - \frac{1}{4}
\frac{\partial^2}{\partial \alpha_r \partial \alpha_s^*} \right) W.
\end{eqnarray}
\end{widetext}
This Wigner function evolution is exactly equivalent to the von Neumann
equation, Eq.~(\ref{Von Neumann equation}).

\subsubsection{Wigner truncation}

Equations of motion for the Wigner function of the form given by
Eq.~(\ref{Wigner function evolution full}) are well known, particularly in
quantum optics \cite{Gardiner2000a}, and even in the case of a few variables
are not easy to solve numerically. The \emph{Wigner truncation}, in which the
third-order derivative terms are dropped, has often been made, since the
resulting equation of motion, having only first-order derivatives on the 
right, is of the form of a Liouville equation. It thus describes an ensemble 
of trajectories obeying an equation of motion which is essentially classical, 
and which can be simulated relatively straightforwardly.

The justification for this truncation is intuitively reasonable; if the quantum
state of the system is such that it is ``almost classical'', then the classical
equation should prevail. One can  present scaling arguments, in which it is
assumed that the Wigner function  behaves like a sharply peaked probability
distribution centered around a  \emph{macroscopic} mean value of the parameters
$ \alpha_j$. These arguments  can then be used to show that the contribution
from the third-order  derivative terms is negligible provided the mean values
of \emph{all} of the $\alpha_j$ are large. This kind of argument has been made
relatively rigorously by Polkovnikov \cite{Polkovnikov2003a}, who has shown how
the third-order derivative terms give a correction to the classical
trajectories.

In the case of colliding condensates this criterion is not valid. The initial 
state of the system is only highly occupied in the modes in the vicinity of the
two incoming momenta, and we want to consider the evolution into a large number
of initially unoccupied modes. However, experience in quantum optics shows that
the Wigner truncation can be valid for such a system; the prime example is the
\emph{degenerate parametric oscillator}, in which two modes of the
electromagnetic field are made to interact by means of a nonlinear crystal to
give a Hamiltonian of the form
\begin{equation}
\label{opo1}
\hat{H}_{\rm dpo} = \hbar \omega \hat{a}^{\dagger} \hat{a}
+ 2 \hbar \omega \hat{b}^{\dagger} \hat{b}
+ g \left\{ \hat{b} \left(\hat{a}^{\dagger}\right)^2
+ \hat{b}^{\dagger} \hat{a}^2 \right\}.
\end{equation}
Here $\hat{a}$ and $\hat{b}$ are destruction operators for field modes with
frequencies $\omega$ and $2\omega$. A Wigner function treatment soon reveals an
equation of motion with first- and third-order derivatives with respect to
corresponding Wigner function variables $\alpha,\alpha^*,\beta,\beta^*$.

In the relevant physical situation, the mode $\hat{a}$ is initially unpopulated
and the mode $\hat{b}$ is derived from an intense laser, and as such is
essentially a classical field --- thus one makes the replacement
\begin{equation}
\label{opo2}
\hat{b} \rightarrow  B e^{-2i\omega t},    
\end{equation}
where $B$ is a classical amplitude, and thus
\begin{eqnarray}
\label{opo3}
\hat{H}_{\rm dpo} \rightarrow \hat{H}'_{\rm dpo} & = &
\hbar \omega \hat{a}^{\dagger} \hat{a} \nonumber \\
&& \hspace{-1cm} \mbox{}
+ g \left\{ B e^{-2i\omega t} \left(\hat{a}^{\dagger} \right)^2 +
B^* e^{2i\omega t} \hat{a}^2\right\}.
\end{eqnarray}
This approximate Hamiltonian produces a Wigner function equation of motion with
only first-order derivatives with respect to $\alpha,\alpha^*$, and is well
verified to give accurate physical predictions for the production of quanta in
the mode $\hat{a}$.  

Alternatively, one could simply drop the third-order terms in the Wigner 
function equation of motion, and one would get equivalent results in the limit 
that one could neglect depletion of the field $\hat{b}$.  This is in spite of
the fact that the occupation of the mode $\hat{a}$ is not large --- thus it
appears that it is sufficient for only some of the modes to be highly occupied
for it to be valid to drop the third-order derivative terms.

\subsubsection{Validity of the Wigner truncation for colliding condensates}
\label{validity}
The case of colliding condensates is analogous to that just discussed. There
are a number of highly occupied modes---those corresponding to the original
condensate packets, and a larger number of unoccupied modes. In the following
we give a heuristic analysis of why the Wigner truncation should be acceptable
for this system.

Consider a multimode Wigner function, which at some time $\tau$ is factorizable
into single-mode functions, of the form
\begin{equation}
\label{Wigner function factorisable}
W \left( \left\{ \alpha_j, \alpha_j^* \right\}, \tau \right) =
\prod_{j \in L} \frac{\Gamma_j}{\pi}
\exp \left[ - \Gamma_j \left| \alpha_j - \alpha_{j_0} \right|^2 \right],
\end{equation}
where $\alpha_{j_0}$ gives the expectation value (coherent) amplitude of the
$j$th mode, and $\Gamma_j \leq 2$ is the inverse width of the Wigner function.
This Wigner function includes both thermal and coherent (for which $\Gamma_j =
2$) statistics, but does not allow the modes to exist as pure number states or
other more elaborate forms. In fact we use exactly this Wigner function when
constructing the initial states of our simulations, for which we find that the
modes display essentially Gaussian statistics at all times, with minimal
correlations between the modes. Thus we expect the following analysis to be
appropriate for the duration of the collisions considered here.

Substituting the Wigner function given by Eq.~(\ref{Wigner function
factorisable}) into the nonlinear portion of Eq.~(\ref{Wigner function
evolution full}) gives the evolution at time $\tau$ as
\begin{widetext}
\begin{eqnarray}
\label{Wigner evolution evaluated 1}
i\hbar \frac{\partial W}{\partial t}^{\rm (nonlin)} & = &
U_0 \sum_{jrst \in L} \left \langle jr| st \right \rangle
\left[ \Gamma_j \left( \alpha_j^* - \alpha_{j_0}^* \right) \alpha_t
- \Gamma_t \alpha_j^* \left( \alpha_t - \alpha_{t_0} \right) \right] \nonumber
\\
&& \hspace{2cm} \times
\left\{ \left[ \alpha_r^* \alpha_s - \delta_{r,s} \right]
- \left[ \frac{\Gamma_r \Gamma_s}{4} \left( \alpha_r^* -
\alpha_{r_0}^* \right) \left( \alpha_s - \alpha_{s_0} \right) -
\frac{\Gamma_r}{2} \delta_{r,s} \right] \right\} W .
\end{eqnarray}
\end{widetext}
Here the first bracketed set of terms within the braces arise from the
first-order derivatives, while the second bracketed set of terms arise from the
third-order derivatives.

Analogously to the definition of the restricted basis field operator, as given
by Eq.~(\ref{Restricted field operator}), we define the classical wavefunction
\begin{equation}
\label{Restricted basis wavefunction}
\Psi_{\mathcal{P}} \left( {\bf x} \right) \equiv \sum_{j \in L} \psi_{j} \left(
{\bf x} \right) \alpha_j,
\end{equation}
which represents a possible state of the \emph{total} restricted basis field,
including both the condensate and noncondensate particles, at any given time.
We also define the related wavefunction
\begin{equation}
\label{Truncation wavefunction 1}
\xi_{\mathcal{P}} \left( {\bf x} \right) \equiv \sum_{j \in L}
\psi_j \left( {\bf x} \right) \frac{\Gamma_j}{2} \alpha_j,
\end{equation}
whose expectation value, calculated using the Wigner function given by
Eq.~(\ref{Wigner function factorisable}), is found to be
\begin{equation}
\label{Truncation wavefunction 2}
\xi_{\mathcal{P}_0} \left( {\bf x} \right) \equiv \left \langle
\xi_{\mathcal{P}} \left( {\bf x} \right) \right \rangle_W =
\sum_{j \in L} \psi_j \left( {\bf x} \right) \frac{\Gamma_j}{2} \alpha_{j_0}.
\end{equation}
Using these wavefunction forms in the Wigner function evolution,
Eq.~(\ref{Wigner evolution evaluated 1}), gives
\begin{widetext}
\begin{eqnarray}
\label{Wigner evolution evaluated 2}
i\hbar \frac{\partial W}{\partial t} ^{\rm (nonlin)} & = &
2U_0 \int d{\bf x} \left[
\left( \xi_{\mathcal{P}}^* - \xi_{\mathcal{P}_0}^* \right) \Psi_{\mathcal{P}} -
\Psi_{\mathcal{P}}^* \left( \xi_{\mathcal{P}} - \xi_{\mathcal{P}_0} \right)
\right] \nonumber \\
&& \hspace{3cm}
\times \left\{ \left[ \left| \Psi_{\mathcal{P}} \right|^2
- \sum_{j \in L} \left| \psi_j \right|^2 \right]
- \left[ \left| \xi_{\mathcal{P}} - \xi_{\mathcal{P}_0} \right|^2 -
\sum_{j \in L} \frac{\Gamma_j}{2} \left| \psi_j \right|^2 \right] \right\} W.
\end{eqnarray}
\end{widetext}
where we have suppressed the explicit spatial dependences and have retained the
ordering of Eq.~(\ref{Wigner evolution evaluated 1}).

To justify the Wigner truncation, we now show that the terms arising from the
cubic derivatives in Eq.~(\ref{Wigner evolution evaluated 2}) are small
compared with the first order derivative terms for all points ${\bf x}$ on the
coordinate space field. This local analysis requires that the inequality
\begin{equation}
\label{Truncation validity 1}
\left| \frac{\left| \xi_{\mathcal{P}} - \xi_{\mathcal{P}_0} \right|^2 -
\sum_{j \in L} \frac{\Gamma_j}{2} \left| \psi_j \right|^2}{
\left| \Psi_{\mathcal{P}} \right|^2 - \sum_{j \in L} \left| \psi_j \right|^2}
\right| \ll 1,
\end{equation}
should hold over all space. A useful quantitative description of the inequality
is in terms of the distributional averages (classical expectation values), for
which we find that
\begin{equation}
\label{Truncation wavefunction expectation}
\left \langle \left| \xi_{\mathcal{P}} - \xi_{\mathcal{P}_0} \right|^2
\right \rangle_W = \sum_{j \in L} \frac{\Gamma_j}{4} \left| \psi_j \right|^2,
\end{equation}
which is of similar magnitude to $\delta_{\mathcal{P}} \left( {\bf x,x}
\right)$, Eq.~(\ref{Restricted delta function}). For the expectation value of
the $\left| \Psi_{\mathcal{P}} \right|^2$ however, a more useful representation
can be obtained using the correspondence of Wigner function averages to the
quantum expectation values, Eq.~(\ref{wig2}). We find that in the general case
(\emph{i.e.} irrespective of the particular form of the Wigner function)
\begin{equation}
\label{Expectation |psi|^2}
\left \langle \left| \Psi_{\mathcal{P}} \left( {\bf x} \right) \right|^2 \right
\rangle_W = n \left( {\bf x} \right) + \frac{1}{2} \delta_{\mathcal{P}} \left(
{\bf x,x} \right),
\end{equation}
where the total density of real particles $n \left( {\bf x} \right)$ is defined
using the restricted basis field operators by
\begin{equation}
\label{Total particle density}
n \left( {\bf x} \right) \equiv \left \langle \hat{n} \left( {\bf x} \right)
\right \rangle =
\left \langle \hat{\Psi}_{\mathcal{P}}^{\dagger} \left( {\bf x} \right)
\hat{\Psi}_{\mathcal{P}} \left( {\bf x} \right) \right \rangle.
\end{equation}
Using Eqs.~(\ref{Truncation wavefunction expectation},\ref{Expectation
|psi|^2}), the validity criterion for the Wigner truncation,
Eq.~(\ref{Truncation validity 1}), becomes
\begin{equation}
\label{Truncation validity 2}
\left| n \left( {\bf x} \right) - \frac{1}{2} \delta_{\mathcal{P}} \left(
{\bf x,x} \right) \right| \gg
\sum_{j \in L} \frac{\Gamma_j}{4} \left| \psi_j \left( {\bf x} \right)
\right|^2.
\end{equation}
Thus in order to justify the truncation it is required that the real particle
density is large compared with the commutator of the restricted field,
$\delta_{\mathcal{P}} \left( {\bf x,x} \right)$.

For a zero-temperature homogeneous field, such that $\Gamma_j=2$ for all modes,
the validity condition becomes simply $N\gg M$, which is similar to that given
by Sinatra \emph{et al.} \cite{Sinatra2002a}. However, for the inhomogeneous
finite-temperature case, the localized truncation condition given by
Eq.~(\ref{Truncation validity 2}) is less easy to justify, especially when one
considers that for these inhomogeneous fields there may be regions where the
total particle density goes to zero. However, the part of the Wigner function
evolution dependent upon interparticle scattering is significant in only those
regions where there is a high particle density, \emph{i.e.} the regions where
the truncation is accurate, so that the truncation can be made over all space
without adversely affecting the accuracy of the approximation. Interestingly,
this justification relies on the relative magnitude of the particle and
mode-function \emph{densities}, rather than the \emph{numbers} of particles and
modefunctions. Thus it appears to be possible to accurately apply the Wigner
truncation to systems in which there are significantly more basis modes than
real particles.

\ 
\subsubsection{Critique of the Validity Criterion}\label{critique}
The validity criterion in the form Eq.~(\ref{Truncation validity 2}), or in
the form for a homogeneous system $N\gg M$, shows that for a given
number of real particles in the system, accurate results will not
result if the number of basis modes $M$ is \emph{too large}.  This is
fundamental to the truncated Wigner function method, in which vacuum
noise is added to every mode.  Methods based on the
P-function\cite{Deuar2006a,Deuar2005a,Gardiner2000a} do not have this
problem, since noise is added dynamically and only to modes with real
occupation. However, as noted in the introduction, other technical
difficulties so far make these methods more difficult to use in
practice.
\subsubsection{Truncated Wigner function Fokker-Planck equation}
Within the validity regime of the Wigner truncation the Wigner function
evolution given by Eq.~(\ref{Wigner function evolution full}) is well
approximated by the \emph{truncated Wigner function Fokker-Planck equation}
\begin{widetext}
\begin{eqnarray}
\label{Truncated Wigner FPE}
i \hbar \frac{\partial W}{\partial t} & \approx & - \sum_{j \in L}
\frac{\partial}{\partial \alpha_j} \left\{ \hbar \omega_j \alpha_j
+ U_0 \sum_{rst \in L} \left \langle jr | st \right \rangle
\alpha_r^* \alpha_s \alpha_t \right\} W \nonumber \\
&& \mbox{} + \sum_{j \in L}
\frac{\partial}{\partial \alpha_j^*} \left\{ \hbar \omega_j \alpha_j^*
+ U_0 \sum_{rst \in L} \left \langle jr | st \right
\rangle^* \alpha_r \alpha_s^* \alpha_t^* \right\} W.
\end{eqnarray}
\end{widetext}
Note that we have also removed from Eq.~(\ref{Truncated Wigner FPE}) the
nonlinear evolution dependent upon $\delta_{r,s}$ (see Eq.~(\ref{Wigner
function evolution full})), which corresponds to terms dependent upon
$\delta_{\mathcal{P}} \left( {\bf x,x} \right)$ in coordinate space,
and whose influence on the total evolution, as we discussed above, is negligible.
Indeed, in order to preserve energy conservation these terms \emph{must} be
removed alongside the cubic derivative terms.

\subsection{Ensemble differential equations}

The Liouville equation, Eq.~(\ref{Truncated Wigner FPE}), gives the equation of
motion for the ensemble of trajectories of the variables $\left\{ \alpha_j,
\alpha_j^* \right\}$. The corresponding equations of motion for individual
trajectories of the system are given by \cite{Gardiner2003b}
\begin{equation}
\label{Mode space SDE}
i\hbar \frac{d \alpha_j}{dt} = \hbar \omega_j \alpha_j
+ U_0 \sum_{rst \in L} \left \langle jr | st \right \rangle
\alpha_r^* \alpha_s \alpha_t,
\end{equation}
where $j \in L$. Every distinct realization of the differential equation uses a
different initial noise field, and hence yields a distinct trajectory in the
phase space of the system --- we describe appropriate initial states in section
\ref{SSSection: initial states}.

By using our previously defined restricted basis wavefunction,
Eq.~(\ref{Restricted basis wavefunction}), in the time-dependent form
$\Psi_{\mathcal{P}} \left( {\bf x},t \right)$, we can rewrite the evolution of
the $j$th low-energy mode amplitude as
\begin{equation}
\label{Hybrid SDE}
i\hbar \frac{d \alpha_j}{dt} = \hbar \omega_j \alpha_j + U_0 \int d{\bf x} \,
\psi_j^* \left| \Psi_{\mathcal{P}} \right|^2 \Psi_{\mathcal{P}},
\end{equation}
which we find to be convenient for numerical simulation. We use
Eq.~(\ref{Hybrid SDE}) (in a dimensionless form) to calculate the central
results of this paper.

Note that although our differential equations are not \emph{stochastic}
differential equations, they do have a stochastic nature that arises from the
random component of the initial field.

\subsubsection{Projected form of the differential equations}

It is instructive to obtain the differential equation for the entire
restricted field, rather than the evolutions of the individual mode amplitudes.
Again using the definition of the restricted basis wavefunction, we find that
Eq.~(\ref{Hybrid SDE}) gives rise to 
\begin{equation}
\label{Coordinate space SDE}
i\hbar \frac{\partial \Psi_{\mathcal{P}}}{\partial t} =
\left[ -\frac{\hbar^2 \nabla^2}{2m} + U_{\rm ext} \right] \Psi_{\mathcal{P}}
+ U_0 \mathcal{P} \left\{ \left[ \left| \Psi_{\mathcal{P}} \right|^2 \right]
\Psi_{\mathcal{P}} \right\},
\end{equation}
where we have replaced the basis eigenenergies with the corresponding diagonal
Hamiltonian and have recognized the form of the projector onto the low-energy
mode space, Eq.~(\ref{Spatial P operator}).

It is straightforwardly shown from any of Eqs.~(\ref{Mode space
SDE},\ref{Hybrid SDE},\ref{Coordinate space SDE}) that the normalization of the
total field for a single trajectory, defined as
\begin{equation}
\label{Field normalisation}
\mathcal{N} \equiv \sum_{j \in L} \left| \alpha_j \right|^2 = \int d{\bf x}
\left| \Psi_{\mathcal{P}} \right|^2,
\end{equation}
is strictly conserved, so that $d \mathcal{N}/ dt = 0$, as is the total field
energy
\begin{equation}
\mathcal{E} = \sum_{j \in L} \hbar \omega_j \left| \alpha_j \right|^2
+ \frac{U_0}{2} \int d{\bf x} \, \left| \Psi_{\mathcal{P}} \right|^4.
\end{equation}
Note that $\mathcal{N}$ and $\mathcal{E}$ are \emph{not} the physically
observable field population and energy, as they include contributions from the
virtual particles (the noise). It could be imagined that the projector serves
to remove those quanta that are shifted outside the low-energy subspace by
pairwise collisions, reducing the total population of the field over time. This
view is incorrect. Rather the projector simply disallows these processes, a
result that is most easily seen from Eq.~(\ref{Mode space SDE}).

\subsubsection{Initial states}
\label{SSSection: initial states}

As the ensemble differential equations contain no dynamic noise sources, to
sample the evolution of the Wigner function we are obliged only to ensure that
the ensemble of \emph{initial states} yields the appropriate Wigner function.
For the zero-temperature collisions presented here we assume an initial Wigner
function of the form given by Eq.~(\ref{Wigner function factorisable}), where
the modes are assumed to display uniformly coherent statistics, for
which
$\Gamma_j = 2$ for all $j$. The corresponding initial state of the $j$th mode
amplitude for a single trajectory is given by
\begin{equation}
\label{Mode initial state}
\alpha_j \left( 0 \right) = \alpha_{j_0} \left( 0 \right)
+ \frac{1}{\sqrt{2 \Gamma_j}} \left( A_j + iB_j \right).
\end{equation}
Here, as before, $\alpha_{j_0}$ is the coherent amplitude of the mode, and can
therefore be identified as the condensate amplitude for the $j$th mode. The
quantities $A_j,B_j$ are real independent Gaussian random variables of zero
mean and unit variance, such that
\begin{eqnarray}
\label{Mode fluctuation properties}
\left \langle A_j \right \rangle = \left \langle B_j \right \rangle =
\left \langle A_i B_j \right \rangle = 0
\nonumber \\
\left \langle A_i A_j \right \rangle =
\left \langle B_i B_j \right \rangle = \delta_{i,j}.
\end{eqnarray}
The equivalent coordinate space initial state is found using
Eqs.~(\ref{Restricted basis wavefunction},\ref{Mode initial state}) to be
\begin{eqnarray}
\Psi_{\mathcal{P}} \left( {\bf x},0 \right) & = &
\sum_{j \in L} \psi_j \left[ \alpha_{j_0} \left( 0 \right)
+ \frac{1}{\sqrt{2\Gamma_j}} \left( A_j + iB_j \right) \right] \\
\label{Initial state 2}
& = &
\Psi_{\mathcal{P}_0} \left( {\bf x},0 \right) + \chi_{\mathcal{P}}
\left( {\bf x} \right).
\end{eqnarray}
Here $\Psi_{\mathcal{P}_0} = \left \langle \Psi_{\mathcal{P}} \right \rangle_W$
is the coherent field amplitude, \emph{i.e.} the condensate wavefunction. The
spatial fluctuations, $\chi_{\mathcal{P}}$ are found from Eq.~(\ref{Mode
fluctuation properties}) to satisfy
\begin{eqnarray}
\left \langle \chi_{\mathcal{P}} \right \rangle
\hspace{-0.1cm} & = & \hspace{-0.1cm} 0, \\
\label{Spatial fluctuations properties 2}
\left \langle \chi_{\mathcal{P}}^{*} \left( {\bf x}' \right)
\chi_{\mathcal{P}} \left( {\bf x} \right) \right \rangle
\hspace{-0.1cm} & = & \hspace{-0.2cm}
\sum_{j \in L} \frac{1}{\Gamma_j} \psi_j^* \left( {\bf x}' \right)
\psi_j \left( {\bf x} \right),
\end{eqnarray}
where Eq.~(\ref{Spatial fluctuations properties 2}) can be seen to be similar
to the restricted delta function defined by Eq.~(\ref{Restricted delta
function}), and proportional to it when all $\Gamma_j = 2$. For our uniformly
coherent initial state we find that the expected total field normalization, as
given by Eq.~(\ref{Field normalisation}), is
\begin{equation}
\left \langle \mathcal{N} \right \rangle = N_0 \left( 0 \right) + \frac{M}{2},
\end{equation}
where
\begin{equation}
N_0 \left( t \right) \equiv \sum_{j \in L} \left| \alpha_{j_0} \left( t \right)
\right|^2 = \int d{\bf x} \left| \Psi_{\mathcal{P}_0} \left( {\bf x},t \right)
\right|^2,
\end{equation}
is the total number of condensate particles and, as before, $M$ is the number
of low-energy modes. Although the number of condensate particles is in general
not conserved, the total number of real particles $N$ is conserved.

\section{Numerical methods}

\subsection{Computational units}

It is convenient for the purposes of numerical simulation to express the
equations in dimensionless computational units. The systems we treat are
initially confined within a harmonic potential
\begin{equation}
U_{\rm harm} \left( {\bf x} \right) = \frac{m}{2} \left( \omega_x^2 x^2 +
\omega_y^2 y^2 + \omega_z^2 z^2 \right),
\end{equation}
which provides
\begin{equation}
\label{Natural units}
x_0 = \sqrt{\frac{\hbar}{2m \omega_x}}, \hspace{0.5cm} t_0 = \frac{1}{\omega_x},
\hspace{0.5cm} \varepsilon_0 = \hbar \omega_x.
\end{equation}
as a natural choice of units for length, time and energy respectively.

In these units Eq.~(\ref{Hybrid SDE}), becomes
\begin{equation}
\label{Hybrid SDE dimensionless}
i \frac{d \alpha_j}{d\tilde{t}} = \tilde{\omega}_j \alpha_j +
\tilde{U}_0 \int d\tilde{\bf x} \,
\tilde{\psi}_j^* \left| \tilde{\Psi}_{\mathcal{P}} \right|^2
\tilde{\Psi}_{\mathcal{P}},
\end{equation}
where we have used tildes to indicate quantities in computational units
($\alpha_j$ is identical in both computational and S.I. units). As an example,
the dimensionless interaction parameter $\tilde{U}_0$ is
\begin{equation}
\label{Nonlinear constant}
\tilde{U}_0 \equiv \frac{U_0}{\varepsilon_0 x_0^3}
= 8 \pi a \sqrt{\frac{2 m \omega_x}{\hbar}}.
\end{equation}

\subsection{Plane wave basis}

While our formalism allows the use of any set of orthonormal basis states
$\psi_j \left( {\bf x} \right)$, the most appropriate choice of basis for
collisions occurring in free space is the plane wave states. We use a
three-dimensional, periodically-bounded rectangular space of volume $V = L_x
\times L_y \times L_z$, for which the (normalized to unity) plane-wave modes
are
\begin{equation}
\psi_j \left( {\bf x} \right) = \frac{1}{\sqrt{V}}
e^{i {\bf k}_j \cdot {\bf x}}.
\end{equation}
Here the wavevector ${\bf k}_j$ associated with the $j$th mode is
\begin{equation}
{\bf k}_j = \frac{2\pi m_j}{L_x} \hat{\bf k}_x
+ \frac{2\pi n_j}{L_y} \hat{\bf k}_y + \frac{2\pi p_j}{L_z} \hat{\bf k}_z,
\end{equation}
where $m_j$, $n_j$ and $p_j$ are integers.

In our dimensionless computational units, the single-particle energy of the
$j$th mode is given by $\tilde{\omega_j} = \tilde{k}_j^2$, and the differential
equation describing the evolution of the $j$th mode, Eq.~(\ref{Hybrid SDE
dimensionless}), becomes
\begin{equation}
\label{Hybrid SDE dimensionless plane wave}
i \frac{d\alpha_j}{d\tilde{t}} = \tilde{k}_j^2 \alpha_j
+ \frac{\tilde{U}_0}{\sqrt{\tilde{V}}} \int d\tilde{\bf x} \,
e^{-i \tilde{\bf k}_j \cdot \tilde{\bf x}} \left| \tilde{\Psi}_{\mathcal{P}}
\right|^2 \tilde{\Psi}_{\mathcal{P}}.
\end{equation}

We have previously defined the low-energy mode subspace $L$ to consist of all
those modes whose energies are less than some cutoff energy $\varepsilon_{\rm
cut}$. It is easily seen that for the plane-wave modes this energy cutoff is
translated to a spherical cutoff in wavevector space. Thus the $L$ subspace
consists of all those modes whose wavevectors satisfy $k_j \leq k_{\rm cut}$,
where the \emph{cutoff wavenumber} is defined (in computational units) as
$\tilde{k}_{\rm cut} \equiv \sqrt{\tilde{\varepsilon}_{\rm cut}}$.

\subsection{Propagation algorithm}
\label{ssection: propagation algorithm}

To propagate the differential equations we employ a modified version of the
\emph{Fourth-order Runge-Kutta in the Interaction Picture} (RK4IP) algorithm
\cite{Ballagh2000a}, which has been used extensively to simulate the
time-dependent \GPE. The only difference in our equations of motion is the
presence of the projector, which is straightforwardly performed in momentum
space, albeit with a small number of required conditions that are worth stating
explicitly.

Using a plane-wave basis, the conversions between coordinate space and mode
space required by Eq.~(\ref{Hybrid SDE dimensionless plane wave}) are achieved
using \emph{Fast Fourier Transform} (FFT) algorithms \cite{Press1992a}, which
require that the two spaces are represented using identical numbers of
rectangularly arranged grid points. Thus, given that the low-energy mode space
is spherically bounded, we must include additional modes (from the $H$
subspace) to form a rectangularly bounded space (the \emph{padded mode-space}).
Using a padded mode space that tangentially bounds $L$ is not adequate, because
with such a grid population scattered from the $L$ subspace into the $H$
subspace, which should be removed by the projector, can be \emph{aliased} by
the FFT back into the $L$ subspace, thereby adversely affecting the accuracy of
the simulation. A rectangularly bounded mode-space grid of extent $4k_{\rm cut}
\times 4k_{\rm cut} \times 4k_{\rm cut}$ is the smallest grid that prevents
aliased components returning to the $L$ subspace, and therefore allows for
accurate calculation of the low-energy subspace dynamics while minimizing
computational memory requirements. This padded mode-space has $\sim 48/\pi$
times as many grid points as there are low-energy modes, and significant
computational memory savings can be made by representing the $L$ subspace on
the full padded grid only when absolutely necessary for the propagation
algorithm.

\section{System of interest}
\label{Section: system of interest}

In our previous letter \cite{Norrie2005a} we presented results obtained from a
single trajectory describing a very similar collision to one that had been
realized experimentally \cite{Ketterle2002a}. However, careful analysis
indicates that the systems presented in \cite{Norrie2005a} may stretch the
validity criteria of the truncated Wigner method, in particular due to an
inappropriately low $k_{\rm cut}$. In this paper we treat colliding systems
that probe a different region of parameter space, one which is both
experimentally possible and well within the validity regime of our formalism.

The condensate part of our initial state (see Eq.~(\ref{Initial state 2})) is
assumed to be composed of two equally populated wavepackets derived from a
single harmonically trapped Bose-Einstein condensate using a short $\pi/2$
Bragg pulse \cite{Phillips1999a} of wavevector $\left( \Delta q,0,0 \right)$,
so that
\begin{equation}
\label{Bragg initial state}
\Psi_{\mathcal{P}_0} \left( {\bf x},t=0 \right) = \frac{1}{\sqrt{2}}
\Psi \left( {\bf x} \right) \left[ e^{+i \frac{\Delta q}{2} x} +
e^{-i \frac{\Delta q}{2} x} \right],
\end{equation}
where the envelope function $\Psi \left( {\bf x} \right)$ is that of the
initial, unscattered condensate wavefunction. We describe this envelope
wavefunction by the $N_0$ atom ground state solution to the time-independent
\GPE. Eq.~(\ref{Bragg initial state}) assumes a centre-of-mass frame, which
reduces the number of low-energy modes required and provides a convenient
symmetry. We assume a zero-temperature initial state, such that the all modes
are initially coherent and $N = N_0 \left( t = 0 \right)$, and we remove the
confining potential at $t = 0$.

We use a (dimensionless) nonlinear parameter of $\tilde{U}_0 = 1\times10^{-2}$,
an initial (total) condensate population of two million, and a relative
wavepacket wavenumber of $\tilde{\Delta q} = 10$. This parameter set
corresponds to a system of $^{23}$Na atoms, initially confined within a
potential with $\omega_x = 2\pi \times 4.57$ Hz, where the wavepackets move
with a relative speed of 4.0 mm\,s$^{-1}$, or alternatively to a $^{87}$Rb system
initially confined within a trap with $\omega_x = 2\pi \times 0.31$ Hz, where
the wavepackets are separating at 0.53 mm\,s$^{-1}$. We assume the trapping
potential is cylindrically symmetric, with $\lambda \equiv \lambda_z /
\lambda_{x,y} = \sqrt{8}$.

For our (dimensionless) parameter set, the Thomas-Fermi chemical potential
\cite{Stringari1999a} is found to be $\tilde{\mu}_{\rm TF} = 21.4$, giving
Thomas-Fermi radii of $\tilde{x}_{\rm TF} = \tilde{y}_{\rm TF} =
2\sqrt{\tilde{\mu}_{\rm TF}} = 9.24$ and $\tilde{z}_{\rm TF} = 2
\sqrt{\tilde{\mu}_{\rm TF}}/\lambda = 3.27$. In order to enclose the colliding
system until packet separation we find that a volume of $\tilde{V} = 48.9
\times 33.5 \times 33.5$ is appropriate.

\subsection{Truncated Wigner method validity criteria}
\label{ssection: Truncated Wigner method validity criteria}
In restricting the system to the low-energy subspace, as described in
section~\ref{sssection: validity of the effective field theory}, it was stated
that the momentum space must be large enough to include all relevant modes to
describe the evolution. For the colliding system this requirement translates to
$k_{\rm cut} > 3\Delta q/2$, such that all possible scattering events directly
involving the initial condensate wavepackets and the halo are included. To meet
this validity criterion we use a low-energy subspace cutoff of $\tilde{k}_{\rm
cut} = 18$, for which $M = 5.4 \times 10^6$. Using a result from
\cite{Morgan1999a} we find that the error in the $s$-wave scattering length in
the effective Hamiltonian, Eq.~(\ref{Effective Hamiltonian 1}), is
approximately 0.5\% for this value of the cutoff.

The second validity criterion for the truncated Wigner method,
Eq.~(\ref{Truncation validity 2}), is most easily expressed as
\begin{equation}
n \left( {\bf x} \right) \gg \delta_{\mathcal{P}} \left( {\bf x,x} \right).
\end{equation}
For a harmonically trapped Thomas-Fermi condensate, the (dimensionless) maximum
density is $\tilde{n} \left( {\bf x} = 0 \right) = \tilde{\mu}_{\rm TF}/
\tilde{U_0}$, so that for our system $\tilde{n} \left( {\bf x} = 0 \right) =
2140$. For a plane wave basis we find that the equiposition restricted delta
function can be expressed as
\begin{equation}
\delta_{\mathcal{P}} \left( {\bf x,x} \right) = \frac{M}{V}
\approx \frac{k_{\rm cut}^3}{6\pi^2}.
\end{equation}
For $\tilde{k}_{\rm cut} = 18$, we find that $\tilde{\delta}_{\mathcal{P}}
\left( {\bf x,x} \right) \approx 100$, so that we expect the Wigner truncation
criterion, Eq.~(\ref{Truncation validity 2}), to be satisfied for this system.

We have simulated (otherwise identical) additional trajectories using a larger
cutoff, $k_{\rm cut} = 20.57$, for which $M = 8.4 \times 10^6$. Using the
extrapolation method outlined in section \ref{sssection: total populations}, we
find that the difference in the calculated total coherent population between
the system with this larger cutoff and our principal system approaches 1.25\%
by the end of the simulation times.

For the remainder of this paper we present all results in SI units, appropriate
for the $^{23}$Na system described above.

\section{Results}

\subsection{Single trajectory results}

\subsubsection{Momentum space}

\begin{figure}
\begin{center}
\includegraphics[width=8.6cm]{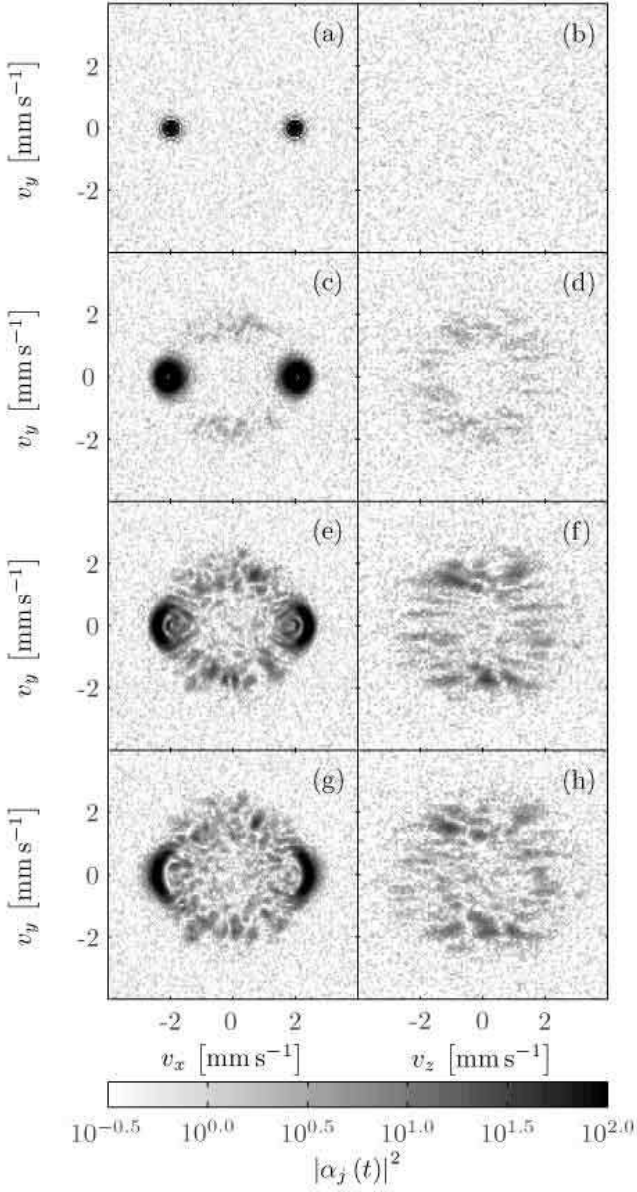}
\end{center}
\vspace{-0.5cm}
\caption{Velocity mode populations on the planes $v_z=0$ (a,c,e,g) and $v_x=0$
(b,d,f,h) at times $t=0$ (a,b), 8.2~ms (c,d), 16.4~ms (e,f) and 37.7~ms (g,h)
for our colliding system described in the text. Note that the simulation volume
exceeds that shown.}
\label{Figure: single run k slices}
\end{figure}

In Fig.~\ref{Figure: single run k slices} we show the mode populations for a
single trajectory of our colliding system, for those modes whose velocities lie
on the planes $v_z = 0$ and $v_x = 0$, at a sequence of times. There are two
major points of interest in this figure. First, we observe from the $v_z = 0$
planes that the original condensate wavepackets broaden and change shape from
ellipsoidal to crescent shaped. Secondly, we observe the generation of a
circular feature, centered at the system centre of mass velocity; this is the
\emph{scattering halo} that is the focus of this work. Note that the mode
population scale of Fig.~\ref{Figure: single run k slices} is logarithmic,
where the scale has been chosen to display the lower populations to best
effect. A consequence of this is that the higher populations saturate for
$\left| \alpha_j \right|^2 \geq 100$, which is rather lower than the population
of $3.2 \times 10^4$ for a mode at the centre of each of the condensate packets
at $t=0$.

We can see from Fig.~\ref{Figure: single run k slices} that the scattering halo
first appears (in our centre of mass frame) centered at a radius slightly less
than the initial central wavenumber of the individual condensate wavepackets,
due to the interaction energy cost associated with creating perturbations in
the field \cite{Bach2002a}. As time progresses the halo broadens, largely
inwards, to occupy (in the average) virtually all those modes within a certain
wavenumber radius. This broadening is a consequence of scattering events
between particles already present in the halo with those in one of the
condensate wavepackets, or other halo particles. Both processes serve to
redistribute population within the halo. In addition we note from
Fig.~\ref{Figure: single run k slices}~(c) that when the halo first appears
there is an angular dependence, with those modes whose polar angles relative to
the collision ($v_x$) axis are closest to $\pi/2$ having the greatest increase
in population.

This anisotropy occurs despite the inherent isotropy of $s$-wave scattering,
because rather than the scattering of single quantum, here we are dealing with
the collective scattering of many particles in the presence of a matter-wave
grating. The probability of any single scattering event is thus proportional to
the overlap integral between the input and output wavefunctions, which means
that modes whose momenta are either perpendicular or parallel to the direction
of the grating have the largest overlap integral and hence experience the
greatest growth.

\begin{figure}
\begin{center}
\includegraphics[width=8.6cm]{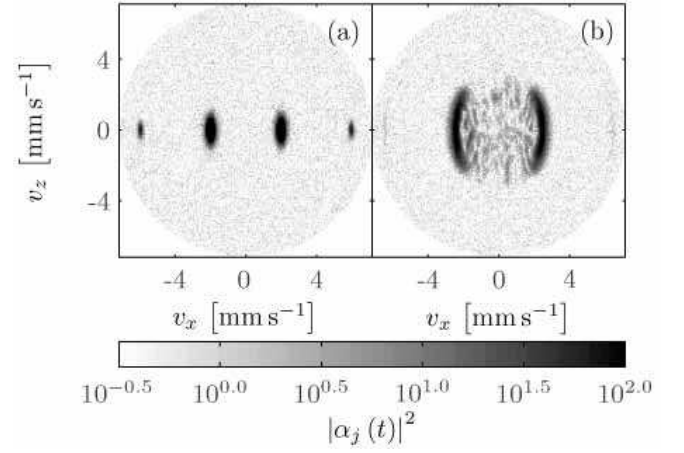}
\end{center}
\vspace{-0.5cm}
\caption{Single trajectory velocity mode populations on the plane $v_y=0$ at
1.6~ms (a) and 37.7~ms (b) for the same collision as in Fig.~\ref{Figure:
single run k slices}. The low-energy subspace boundary is visible as a circle,
beyond which no population occurs.}
\label{Figure: single run higher k}
\end{figure}

Although not visible in Fig.~\ref{Figure: single run k slices}, three-wave
mixing between the condensate packets gives rise to additional wavepackets
centered (in momentum space) at $k_x = \pm 3\Delta q/2$. However, due to the
kinetic energy mismatch involved in their generation, these packets are
transient, and have essentially disappeared by the time the initial wavepackets
are separated. This effect is shown in Fig.~\ref{Figure: single run higher k},
where these higher order packets are present shortly after the collision
begins, and are absent by 37.7~ms. Additionally, this figure shows the
asymmetry in the condensate wavepackets both initially and as a result of
anisotropic broadening due to the oblate nature of the initial trapped
condensate.

\begin{figure}
\begin{center}
\includegraphics[width=8.6cm]{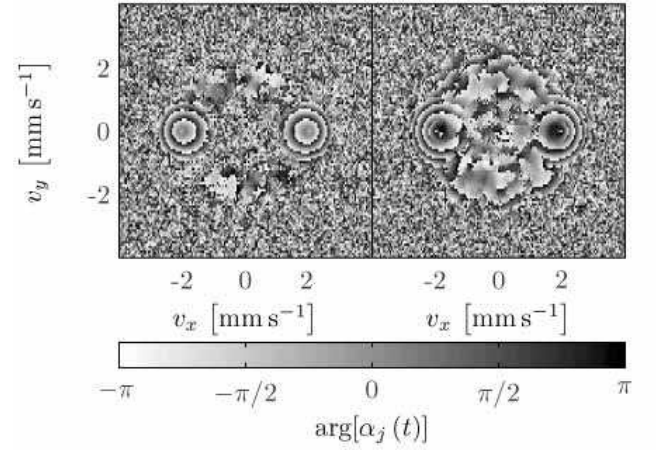}
\end{center}
\vspace{-0.5cm}
\caption{Single trajectory velocity mode phases on the plane $v_z~=~0$ at
8.2~ms (left) and  16.4~ms (right), corresponding to Fig.~\ref{Figure: single
run k slices}~(c,e).}
\label{Figure: single run phase only}
\end{figure}

The scattering halo is not uniformly populated. Rather it is characterized by
distinct patches of high population separated by regions of low population. In
Fig.~\ref{Figure: single run phase only} we plot the phases of the modes on the
plane $v_z = 0$ at times corresponding to the second and third rows of
Fig.~\ref{Figure: single run k slices}. From Fig.~\ref{Figure: single run phase
only}~(left) we observe that as the population in the halo begins to grow,
small regions of relatively constant phase are formed, which by comparing
Figs.~\ref{Figure: single run k slices} and \ref{Figure: single run phase only}
can be identified with those regions of relatively high population. Hence we
label these small, highly-populated regions of momentum space \emph{phase
grains}. The size and aspect ratio of these phase grains is very similar to
those of the initial condensate wavepackets in momentum space, a result that we
explore more quantitatively in section \ref{ssection: autocorrelation
function}. We note that the phase profile of each grain is established prior to
any significant population gain.

\subsubsection{Coordinate space}

\begin{figure}
\begin{center}
\resizebox{8.5cm}{!}{\includegraphics[width=8.6cm]{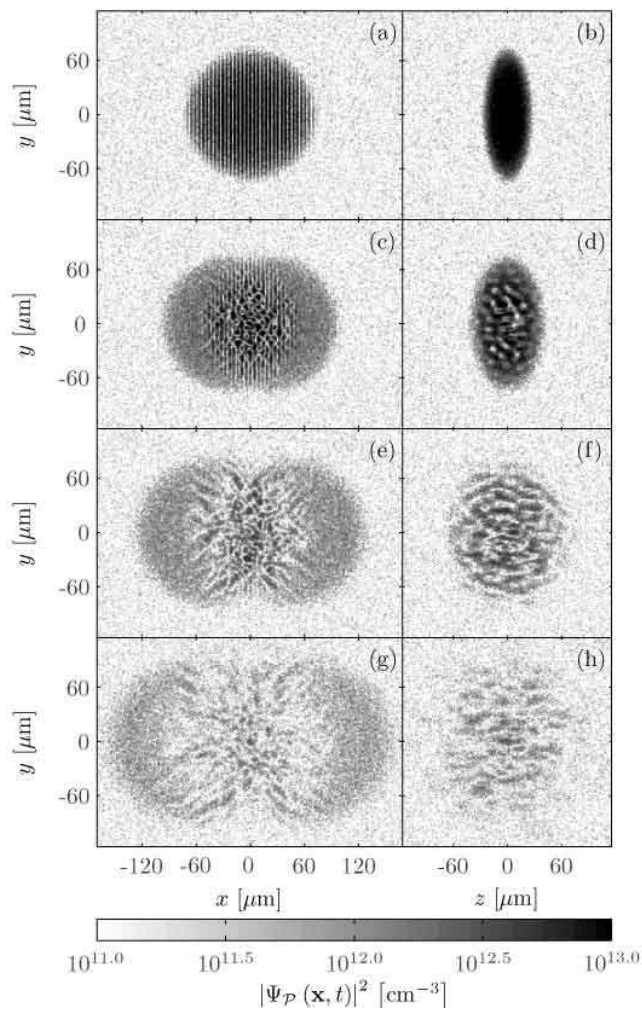}}
\end{center}
\vspace{-0.5cm}
\caption{Coordinate space density for the same collision as Fig.~\ref{Figure:
single run k slices} on the planes $z=0$ (a,c,e,g) and $x=0$ (b,d,f,h) at times
$t = 0$ (a,b), 12.6~ms (c,d), 25.1~ms (e,f) and 37.7~ms (g,h). Note the
logarithmic density scale.}
\label{Figure: single run x slices}
\end{figure}

In Fig.~\ref{Figure: single run x slices} we show the evolving total coordinate
space density $\left| \Psi_{\mathcal{P}} \left( {\bf x},t \right) \right|^2$ on
the planes $z = 0$ and $x = 0$ for the trajectory shown in the preceding
section.  The initial state of the field, shown by Fig.~\ref{Figure: single run
x slices}~(a,b), displays the modulated ground state condensate profile given
by Eq.~(\ref{Bragg initial state}), while the vacuum noise component is
observed to be spatially uniform in its average density, as required for the
plane-wave basis, extending into and distorting the condensate profile.

As the collision proceeds we observe that at the centre of the system, where
the particle density is highest, the regular fringe pattern arising from the
overlapping wavepackets begins to break down, indicating that quanta are being
generated with momenta other than those of the condensate wavepackets. By
25.1~ms into the collision, as shown in Fig.~\ref{Figure: single run x
slices}~(e), the fringes appear to be completely absent. Following this
breakdown, the high density region of the field comprises two distinct types of
wavefunction, a relatively smooth outer portion and a fragmented central
region. In section~\ref{ssection: coherent and incoherent fields}, we identify
the smooth outer shells as condensate that survives the collision, while the
central region forms the scattered halo, together with a very small condensate
remnant. The highest density during the course of the simulation remains close
to the origin, which somewhat counterintuitively leads to the scattered quanta
being more dense than the condensates at later times. After separation of the
condensate packets, scattering events taking place within the central
(fragmented) region involve at most one condensate quantum, and redistribute
population within the halo.

Estimating the separation time using Thomas-Fermi condensate wavepackets of
unchanging shape returns 30.6~ms, significantly longer than that observed in
our simulation. The reduction in separation time reflects the significantly
distorted condensate packets, from initially ellipsoidal to a shape most easily
described as a hemi-ellipsoidal shell (see Fig.~\ref{Figure: single run x
slices}~(g)).

We observe from Fig.~\ref{Figure: single run x slices}~(g) that near the end of
the simulation, the coordinate space field is beginning to resemble the
momentum space field, as shown by Fig.~\ref{Figure: single run k slices}~(g).
By this time the rate of change of the mode populations is very small, and we
therefore expect that Fig.~\ref{Figure: single run k slices}~(g) will give an
excellent indication of the coordinate space density distribution for the
system following a sufficiently long ballistic expansion.

\subsubsection{Turbulence}

\begin{figure}
\begin{center}
\includegraphics[width=8.6cm]{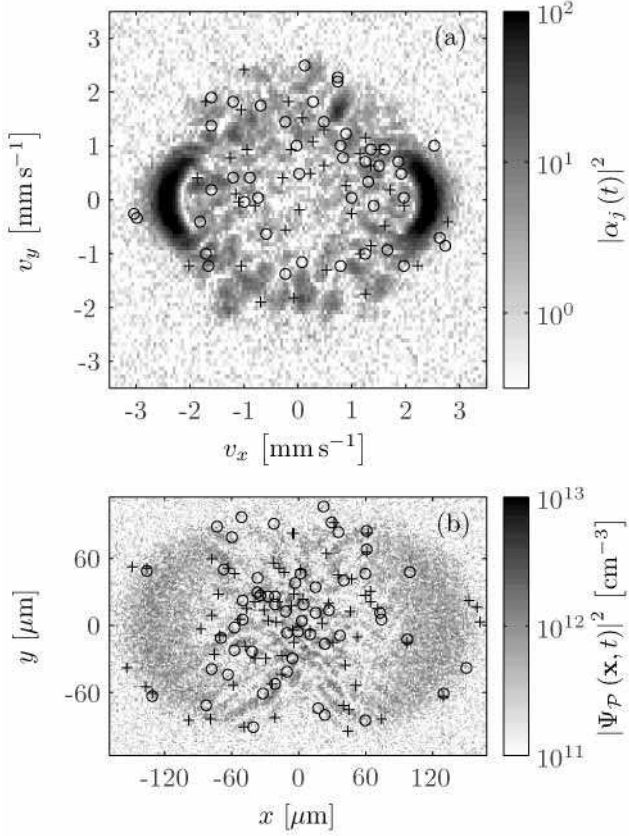}
\end{center}
\vspace{-0.5cm}
\caption{(a) Velocity mode populations on the plane $v_z = 0$ and (b)
coordinate space density on the plane $z = 0$ at 37.7~ms into the collision.
Crosses and circles respectively show unit vortices of positive and negative
sense.}
\label{Figure: vortex both slices}
\end{figure}

Within the scattering halo, a large number of vortices have been detected
between the highly populated grains in both velocity and coordinate spaces. We
demonstrate this in Fig.~\ref{Figure: vortex both slices}, where we show the
velocity mode populations on the plane $v_z = 0$ and the coordinate space
density on the plane $z = 0$ at the end of the collision, together with the
detected vortices. Note that we plot in Fig.~\ref{Figure: vortex both slices}
only those vortices that reside in regions of relatively high average mode
population ($\mathcal{N}_j>1$) or density ($\left| \Psi_{\mathcal{P}}
\right|^2 > 3\times 10^{11}$~cm$^{-3}$), which filters out those vortices that
are present in the field solely by the presence of the virtual particle
fluctuations, and those that are physically observable, although the
threshold choice is somewhat arbitrary.

We have observed (but not plotted) that a significant growth in the number of
physical vortices coincides with the growth of the scattering halo, indicating
that, in much the same way as the halo growth can be viewed as amplification of
the vacuum fluctuations, the observed vortices are amplifications of the
underlying quantum vortices. This provides the basis for our identification of
\emph{quantum turbulence} within the scattering halo.

\subsection{Coherent and incoherent fields}
\label{ssection: coherent and incoherent fields}

We have assembled an ensemble of 80 individual trajectories of our colliding
system, where the initial states of each differ only in the applied virtual
particle noise, as described in section \ref{SSSection: initial states}. In the
following sections we use this ensemble to calculate various quantum statistics
of the colliding system.

\subsubsection{Momentum space}

\begin{figure}
\begin{center}
\includegraphics[width=8.6cm]{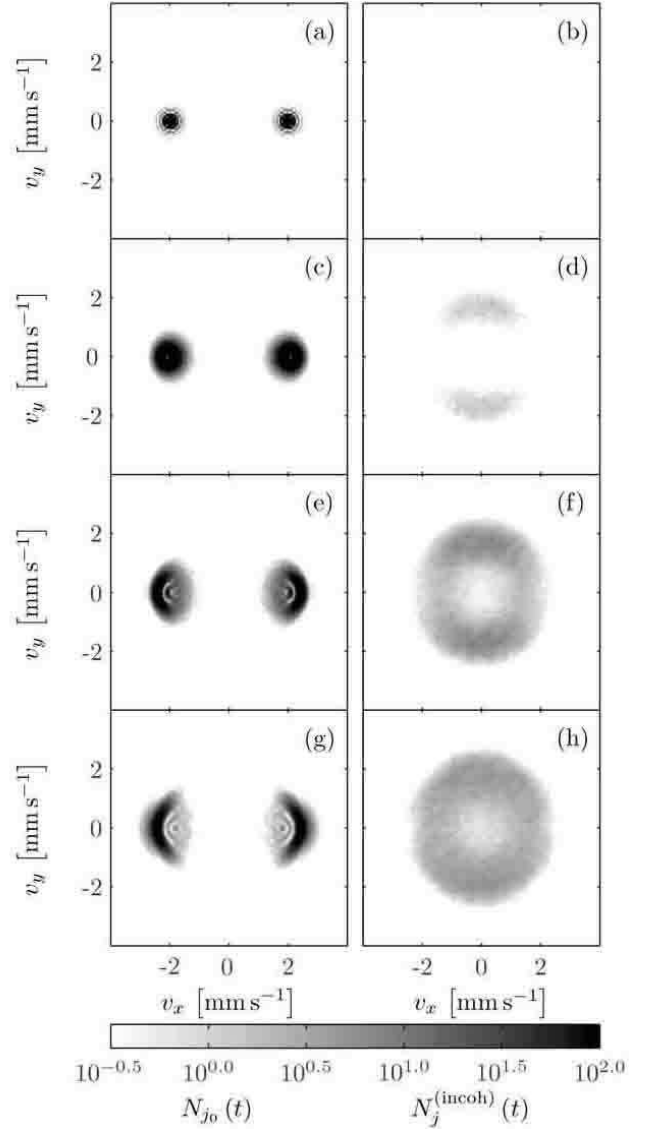}
\end{center}
\vspace{-0.5cm}
\caption{Coherent (a,c,e,g) and incoherent (b,d,f,h) velocity mode populations
on the plane $v_z~=~0$ at times 0 (a,b), 8.2~ms (c,d), 16.4~ms (e,f) and
37.7~ms (g,h).}
\label{Figure: Coh incoh k}
\end{figure}

Using the correspondence between moments of the Wigner function and symmetrized
operator products, Eq.~(\ref{wig2}), we can calculate the \emph{total}
expectation population of the $j$th mode, $N_j \left( t \right)$, using
\begin{equation}
N_j \left( t \right) \equiv
\left \langle \hat{N}_j \right \rangle \left( t \right)
= \left \langle \hat{a}_j^{\dagger} \hat{a}_j \right \rangle \left( t \right)
= \left \langle \left| \alpha_j \left( t \right) \right|^2 \right \rangle_W -
\frac{1}{2}.
\end{equation}
Using the coherent amplitude of the $j$th mode
\begin{equation}
\alpha_{j_0} \left( t \right) \equiv
\left \langle \hat{a}_j \right \rangle \left( t \right)
= \left \langle \alpha_j \left( t \right) \right \rangle_W,
\end{equation}
we can calculate the \emph{coherent} (\emph{i.e.} condensate) part of the total
mode population as
\begin{equation}
N_{j_0} \left( t \right) \equiv \left| \left \langle \alpha_j \left( t \right)
\right \rangle_W \right|^2.
\end{equation}
In Fig.~\ref{Figure: Coh incoh k}~(a,c,e,g) we plot the coherent mode
populations on the plane $v_z = 0$ for our colliding system. From these plots
we observe that the condensate population is restricted to two wavepackets
that, consistent with the behavior shown in Fig.~\ref{Figure: single run k
slices}, broaden and change shape over the course of the collision. A benefit
of plotting the coherent populations only is that the small-scale structure of
the condensate packets is rather more apparent here than in Fig.~\ref{Figure:
single run k slices}. Although not shown in Fig.~\ref{Figure: Coh incoh k}, the
higher order wavepackets, when they appear, are also found to contain
condensate particles.

The corresponding \emph{incoherent} mode populations are calculated as the
difference
\begin{equation}
N_j^{\left( {\rm incoh} \right)} \left( t \right)
\equiv N_j \left( t \right) - N_{j_0} \left( t \right).
\end{equation}
In Fig.~\ref{Figure: Coh incoh k}~(b,d,f,h) we plot the incoherent mode
populations on the plane $v_z = 0$ for the collision. From these plots we
observe that from an initial state with zero incoherent population (as required
by our assumed initial Wigner function), incoherent population builds up
initially as a semi-spherical scattering halo, where those modes occupied by
the condensate wavepackets have a relatively small incoherent population.
Subsequently population is transferred, via further scattering events, to
broaden the distribution of incoherent quanta.

\subsubsection{Coordinate space}

\begin{figure}
\begin{center}
\includegraphics[width=8.6cm]{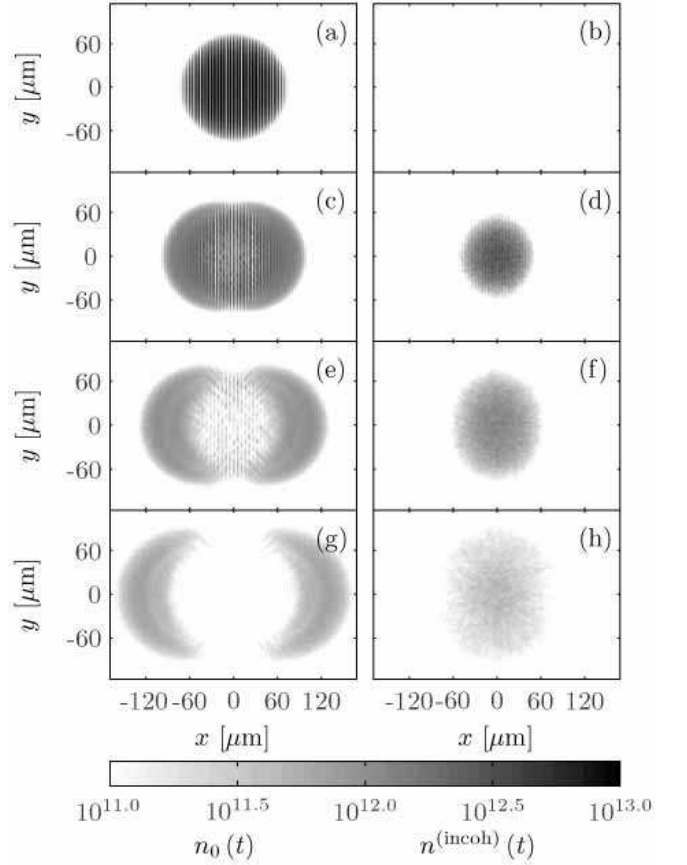}
\end{center}
\vspace{-0.5cm}
\caption{Coherent (a,c,e,g) and incoherent (b,d,f,h) particle densities for
our colliding system on the plane $z~=~0$ at times 0 (a,b), 12.6~ms (c,d),
25.1~ms (e,f) and 37.7~ms (g,h).}
\label{Figure: Coh incoh x}
\end{figure}

The coherent and incoherent fields can also be calculated in coordinate space.
Using the coherent probability amplitude
\begin{equation}
\Psi_{\mathcal{P}_0} \left( {\bf x},t \right) \equiv
\left \langle \hat{\Psi}_{\mathcal{P}} \left( {\bf x} \right) \right \rangle
\left( t \right)
= \left \langle \Psi_{\mathcal{P}} \left( {\bf x},t \right) \right \rangle_W,
\end{equation}
the coherent particle density (condensate density) can be calculated as
\begin{equation}
n_0 \left( {\bf x},t \right) = \left| \Psi_{\mathcal{P}_0} \left( {\bf x},t
\right) \right|^2 = \left| \sum_{j \in L} \psi_j \left( {\bf x} \right)
\alpha_{j_0} \left( t \right) \right|^2.
\end{equation}
The corresponding \emph{incoherent} particle density is found using both the
total particle density, Eqs.~(\ref{Expectation |psi|^2},\ref{Total particle
density}), and the coherent particle density to be
\begin{eqnarray}
n^{\left( {\rm incoh} \right)} \left( {\bf x},t \right) & \equiv &
n \left( {\bf x},t \right) - n_0 \left( {\bf x},t \right) \\
& = &
\left \langle \left| \Psi_{\mathcal{P}} \left( {\bf x},t \right) \right|^2
\right \rangle_W -
\left| \left \langle \Psi_{\mathcal{P}} \left( {\bf x},t \right) \right
\rangle_W \right|^2 \nonumber \\
&& \hspace{1cm} \mbox{}
- \frac{1}{2} \delta_{\mathcal{P}} \left( {\bf x,x} \right),
\end{eqnarray}
where the restricted delta function, $\delta_{\mathcal{P}} \left( {\bf x,x}'
\right)$ is defined by Eq.~(\ref{Restricted delta function}). We plot both the
coherent and incoherent particle densities for our colliding system in
Fig.~\ref{Figure: Coh incoh x}.

From Fig.~\ref{Figure: Coh incoh x}~(a,b) we observe that the initial state is
uniformly coherent, consistent with the assumed initial Wigner function. As the
collision progresses we observe that significant incoherent particle density
develops close to the origin, generated from the condensate particles. Note
that Fig.~\ref{Figure: Coh incoh x}~(e,g) shows that a small interference
pattern exists in this central region even late into the collision, which was
not apparent from a single trajectory (see Fig.~\ref{Figure: single run x
slices}).

\subsubsection{Total coherent and incoherent populations}
\label{sssection: total populations}

One of the important quantities that can be calculated from these simulated
collisions is the total number of particles scattered out of the condensate
wavepackets. In our previous Letter \cite{Norrie2005a} we calculated this
quantity approximately using a spatially dependent counting method in momentum
space. Such methods are routinely used experimentally. However, using our
ensemble we can obtain the total coherent population
\begin{equation}
N_0 \left( t \right) \equiv \sum_{j \in L} \left| \alpha_{j_0} \left( t \right)
\right|^2,
\end{equation}
and incoherent population
\begin{equation}
N^{\left( {\rm incoh} \right)} \left( t \right) \equiv N - N_0 \left( t \right),
\end{equation}
with a great deal more accuracy.

A well-known result for averages of randomly distributed variables is that the
convergence of a finite number of samples $\mathcal{M}$ to the true value
scales as $1/\sqrt{\mathcal{M}}$ (where convergence is measured as the error in
the calculated average). We define the total coherent field population,
calculated from $\mathcal{M}$ trajectories, as
\begin{equation}
\label{Total coherent pop M runs}
N_0 \left( t \right)^{\left( \mathcal{M} \right)} \equiv
\sum_{j \in L} \left| \frac{1}{\mathcal{M}} \sum_{m = 1}^{\mathcal{M}}
\alpha_j \left( t \right)^{\left( m \right)} \right|^2,
\end{equation}
where $\alpha_j \left( t \right)^{\left( m \right)}$ is the time-dependent mode
amplitude of the $m$th trajectory. Given the large number of modes within our
colliding system, and the magnitude squaring operation in Eq.~(\ref{Total
coherent pop M runs}), it is expected that
\begin{equation}
\label{Coherent pop convergence}
N_0 \left( t \right)^{\left( \mathcal{M} \right)} \approx N_0 \left( t \right) +
\frac{C \left( t \right)}{\mathcal{M}},
\end{equation}
where $N_0 \left( t \right) = N_0 \left( t \right)^{\left( \infty \right)}$.
Here $C \left( t \right)$ is some dimensionless time-dependent function,
independent of $\mathcal{M}$, whose exact form is in fact unimportant for our
purposes. Assuming that Eq.~(\ref{Coherent pop convergence}) holds, we can
therefore use any two calculated $N_0 \left( t \right)^{\left( \mathcal{M}
\right)}$ of dissimilar $\mathcal{M}$ to obtain the ensemble limit. In
particular, using $\mathcal{M} = 1$ for one of these two, we find that
\begin{equation}
N_0 \left( t \right) \approx \frac{N_0 \left( t \right)^{\left( 1 \right)} -
\mathcal{M} N_0 \left( t \right)^{\mathcal{M}}}{1 - \mathcal{M}}.
\end{equation}
In practise we find that this extrapolation method is extremely accurate using
just \emph{two} trajectories. This result makes a survey of the parameter
dependence of the global field quantities numerically efficient although,
unfortunately, an equivalent extrapolation method for local field quantities is
significantly less accurate.

\begin{figure}
\begin{center}
\resizebox{8.5cm}{!}{\includegraphics[width=8.6cm]{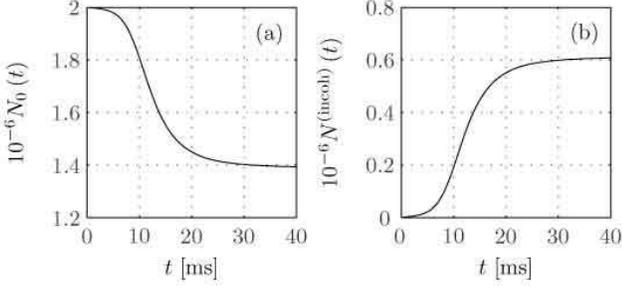}}
\end{center}
\vspace{-0.5cm}
\caption{Total coherent (a) and incoherent (b) populations for the colliding
system calculated using the extrapolation method outlined in the text.}
\label{Figure: Coh incoh populations}
\end{figure}

In Fig.~\ref{Figure: Coh incoh populations} we plot the total coherent and
incoherent populations for our colliding system, calculated using our
extrapolation method with $\mathcal{M} = 80$. From these curves we
observe that the coherent population monotonically decreases with time, with
the most significant depletion occurring between 5~ms and 25~ms into the
collision, by which time 30\% of the total population is incoherent. Note that
the incoherent population continues to increase even after separation of the
condensate wavepackets, which results from decohering interactions involving
individual wavepackets only. Note also that the difference in the calculated
populations is less than 1.5\% for all $\mathcal{M} \in \left[ 2,80 \right]$.

\subsection{Local correlation functions}

\subsubsection{Momentum space}

The normalized second-order equiposition mode-space correlation function
\begin{equation}
\label{g22j quantum}
g_j^{\left( 2 \right)} \left( t \right) \equiv \frac{\left \langle
\hat{a}_j^{\dagger} \hat{a}_j^{\dagger} \hat{a}_j \hat{a}_j \right \rangle
\left( t \right)}{\left[ \left \langle \hat{a}_j^{\dagger} \hat{a}_j
\right \rangle \left( t \right) \right]^2},
\end{equation}
probes the quantum statistics of the $j$th momentum mode. If a mode displays
coherent statistics, then we should observe that $g_j^{\left( 2 \right)} = 1$.
Conversely, if the mode displays Gaussian (\emph{i.e.} thermal) statistics,
then we should find that $g_j^{\left( 2 \right)} = 2$. By applying the
correspondence of the Wigner function moments to the symmetrized quantum
expectation values, Eq.~(\ref{wig2}), we find that $g_j^{\left( 2 \right)}
\left( t \right)$ can be calculated using
\begin{equation}
\label{g22j Wigner}
g_j^{\left( 2 \right)} \left( t \right) = \frac{
\left \langle \left| \alpha_j \left( t \right) \right|^4 \right \rangle_W
- 2 \left \langle \left| \alpha_j \left( t \right) \right|^2 \right \rangle_W
+ \frac{1}{2}}{\left[ \left \langle \left| \alpha_j \left( t \right) \right|^2
\right \rangle_W - \frac{1}{2} \right]^2}.
\end{equation}

\begin{figure}
\begin{center}
\includegraphics[width=8.6cm]{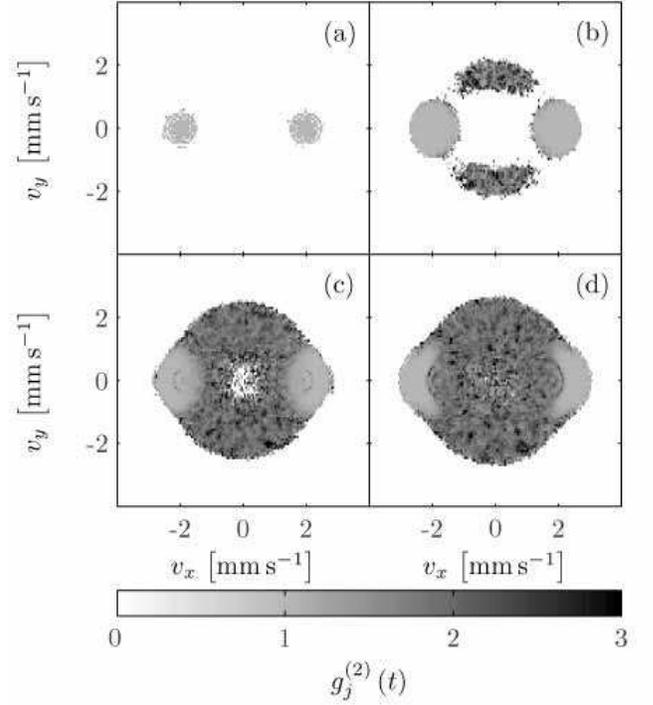}
\end{center}
\vspace{-0.5cm}
\caption{Normalized second-order mode space correlation function $g^{\left( 2
\right)}_j \left( t \right)$ for the colliding system, on the plane $v_z~=~0$,
at times 0 (a), 8.2~ms (b), 16.4~ms (c) and 37.7~ms (d). Results are only shown
for those modes with $N_j > 1/2$.}
\label{Figure: g22 k}
\end{figure}

We plot $g_j^{\left( 2 \right)} \left( t \right)$ on the plane $v_z = 0$ for
our colliding system in Fig.~\ref{Figure: g22 k}. Due to the normalized
character of the correlation function and the finite number of ensemble
members, those modes with small populations give highly noisy results.
Therefore we have plotted in Fig.~\ref{Figure: g22 k} only those modes whose
real particle population is larger than one half. (Of course the quantum
statistics of a mode are only defined when that mode is populated.) From
Fig.~\ref{Figure: g22 k} we observe that the condensate wavepackets have
$g_j^{\left( 2 \right)} = 1$, and are therefore coherently populated, whereas
the halo modes have $g_j^{\left( 2 \right)} \approx 2$, characteristic of
Gaussian statistics. The higher level of noise in the results returned for the
halo modes as opposed to the condensate modes is a consequence of the much
lower average population of the halo modes, and would be reduced for an
increased number of ensemble members.

\subsubsection{Coordinate space}

We can similarly characterize the quantum statistics of the system in
coordinate space, for which the appropriate analogue to Eq.~(\ref{g22j
quantum}) is
\begin{equation}
g^{\left( 2 \right)} \left( {\bf x},t \right) \equiv
\frac{\left \langle \hat{\Psi}_{\mathcal{P}}^{\dagger} \left( {\bf x} \right)
\hat{\Psi}_{\mathcal{P}}^{\dagger} \left( {\bf x} \right)
\hat{\Psi}_{\mathcal{P}} \left( {\bf x} \right)
\hat{\Psi}_{\mathcal{P}} \left( {\bf x} \right) \right \rangle \left( t
\right)}{\left[ \left \langle \hat{\Psi}_{\mathcal{P}}^{\dagger}
\left( {\bf x} \right) \hat{\Psi}_{\mathcal{P}} \left( {\bf x} \right) \right
\rangle \left( t \right) \right]^2}.
\end{equation}
By expanding the field operators on the restricted mode basis, Eq.~(\ref{Full
field operator}) and applying the Wigner function moment correspondences,
Eq.~(\ref{wig2}), we find that the second order normalized coordinate space
correlation function can be calculated using
\begin{equation}
\label{g22 x Wigner}
g^{\left( 2 \right)} \left( {\bf x},t \right) =
\frac{\left \langle \left| \Psi_{\mathcal{P}} \right|^4 \right \rangle_W
- 2 \left \langle \left| \Psi_{\mathcal{P}} \right|^2 \right \rangle_W
\delta_{\mathcal{P}} + \frac{1}{2} \delta_{\mathcal{P}}^2}{\left[ \left
\langle \left| \Psi_{\mathcal{P}} \right|^2 \right \rangle_W
- \frac{1}{2} \delta_{\mathcal{P}} \right]^2},
\end{equation}
where for conciseness we have written $\Psi_{\mathcal{P}} = \Psi_{\mathcal{P}}
\left( {\bf x},t \right)$ and $\delta_{\mathcal{P}} = \delta_{\mathcal{P}}
\left( {\bf x,x} \right)$. As with $g^{\left( 2 \right)}_j \left( t \right)$,
Eq.~(\ref{g22 x Wigner}) should return unity for regions of coherently
distributed density and two for regions of Gaussian distributed density.

\begin{figure}
\begin{center}
\includegraphics[width=8.6cm]{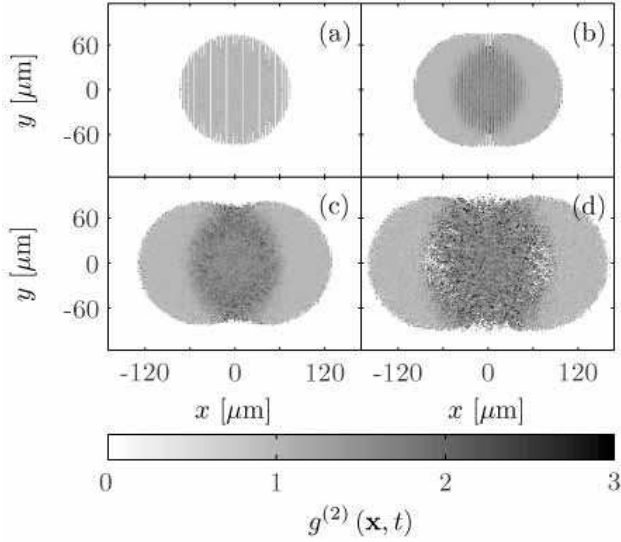}
\end{center}
\vspace{-0.5cm}
\caption{Normalized second-order equiposition coordinate space correlation
function $g^{\left( 2 \right)} \left( {\bf x},t \right)$ for the collision on
the plane $z~=~0$ at times 0 (a), 12.6~ms (b), 25.1~ms (c) and 37.7~ms (d).
Results are only shown for those points where $n \left( {\bf x} \right) > 3
\times 10^{11}$~cm$^{-3}$.}
\label{Figure: g22 x}
\end{figure}

In Fig.~\ref{Figure: g22 x} we plot $g^{\left( 2 \right)} \left( {\bf x},t
\right)$ for the collision, calculated using Eq.~(\ref{g22 x Wigner}), at a
sequence of times. These plots again show a uniformly coherent initial state,
as required, and as time progresses we observe that close to the centre of the
collision the system becomes increasing incoherent, reflecting the creation of
the halo, so that by 37.7~ms the central region shows $g^{\left( 2 \right)}
\left( {\bf x},t \right) \sim 2$.

\subsection{Pair creation correlation functions}

The halo formation process can be viewed as a four-wave mixing, in which the
two input particles are from each of the condensate wavepackets, and the output
particles appear (in the centre of mass frame) in modes of approximately (due
to the finite momentum spread of the condensate wavepackets) opposite momentum.
It is expected therefore that modes of opposite momenta will display
correlations in both their amplitude and population, at least at early times.

An appropriate (normalized) correlation function to quantify the amplitude
correlation between modes of opposing momenta is
\begin{equation}
\label{g02 quantum}
g^{\left( 0,2 \right)}_{ij} \left( t \right) \equiv
\frac{\left \langle \hat{a}_i \hat{a}_j \right \rangle \left( t \right) - \left
\langle \hat{a}_i \right \rangle \left( t \right) \left \langle \hat{a}_j \right
\rangle \left( t \right)}{\sqrt{\left \langle \hat{a}^{\dagger}_i \hat{a}_i
\right \rangle \left( t \right) \left \langle \hat{a}_j^{\dagger} \hat{a}_j
\right \rangle \left( t \right)}},
\end{equation}
where we use ${\bf k}_j = -{\bf k}_i$ and the second term in the numerator
corrects for the non-zero amplitude expectation value for condensate modes.
This function can be written in terms of moments of the Wigner function as
\begin{equation}
g^{\left( 0,2 \right)}_{ij} \left( t \right) = 
\frac{\left \langle \alpha_i \left( t \right) \alpha_j \left( t \right) \right
\rangle_W - \left \langle \alpha_i \left( t \right) \right \rangle_W \left
\langle \alpha_j \left( t \right) \right \rangle_W}{\sqrt{\left[ \left \langle
\left| \alpha_i \left( t \right) \right|^2 \right \rangle_W - \frac{1}{2}
\right] \left[ \left \langle \left| \alpha_j \left( t \right) \right|^2 \right
\rangle_W - \frac{1}{2} \right]}}.
\end{equation}
We plot the magnitude of $g^{\left( 0,2 \right)}_{ij} \left( t \right)$ for
${\bf k}_j = -{\bf k}_i$ at a sequence of times in Fig.~\ref{Figure: g02 pair
correlation}. (Due to our choice of $i,j$ momenta the results are symmetric
about the origin.) From these slices we observe that there is a definite
amplitude correlation between modes of opposite momenta, largest at early times
($g^{\left( 0,2 \right)}_{ij} \left( t \right)\sim0.3$ at 6.6~ms), and
decreasing as the collision progresses. The degradation in the correlation as
time progresses is a consequence of the subsequent scattering events, which
leads to the correlation between paired modes becoming essentially zero at late
times.

\begin{figure}
\begin{center}
\includegraphics[width=8.6cm]{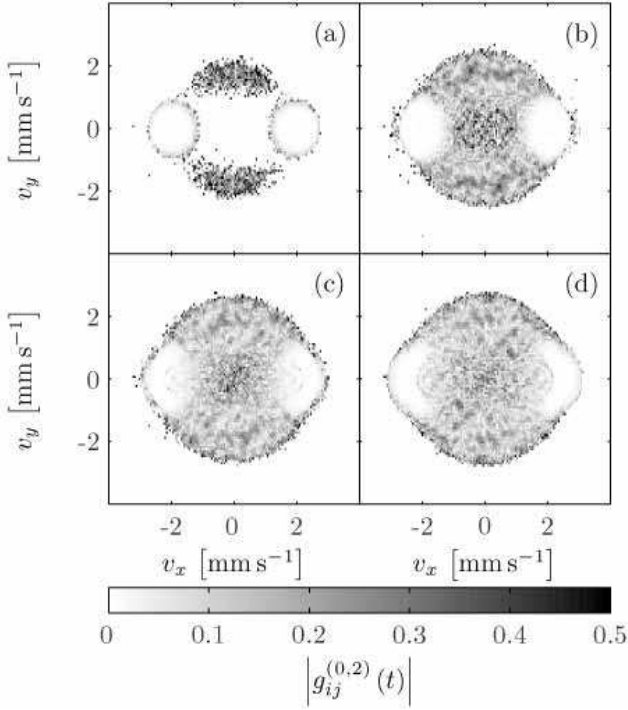}
\end{center}
\vspace{-0.5cm}
\caption{Magnitude of the mode amplitude pair creation correlation
function $g^{\left( 0,2 \right)}_{ij} \left( t \right)$ on
the plane $v_z~=~0$ at 6.6~ms (a), 13.1~ms (b), 19.7~ms (c) and 37.7~ms (d).
Results are only shown for those modes with $N_j > 1/4$.}
\label{Figure: g02 pair correlation}
\end{figure}

Note that in Fig.~\ref{Figure: g02 pair correlation} we have only plotted
results for which $N_j > 1/4$. Although we can calculate the correlation
function for all modes and at all times, the results have no physical meaning
for unpopulated modes. Furthermore, modes with very low population produce very
noisy results.

The analogue to Eq.~(\ref{g02 quantum}) for measuring the population
correlation between paired modes is
\begin{equation}
g^{\left( 2 \right)}_{ij} \left( t \right) \equiv \frac{\left \langle
\hat{a}_i^{\dagger} \hat{a}_j^{\dagger} \hat{a}_i \hat{a}_j \right \rangle
\left( t \right)}{\left \langle \hat{a}^{\dagger}_i \hat{a}_i \right \rangle
\left( t \right) \left \langle \hat{a}_j^{\dagger} \hat{a}_j \right \rangle
\left( t \right)}.
\end{equation}
We have calculated this quantity, and have observed essentially the same
behavior as for the amplitude correlations, \emph{i.e.} a definite correlation
exists at early times, that decreases rapidly with time. We note however that a
very large error exists in the calculations of $g^{\left( 2 \right)}_{ij}$ (of
order $\pm2$ at $t = 6$~ms), much larger than for the amplitude correlations,
so that quantifying the degree of this correlation is difficult.

\subsection{Autocorrelation function}
\label{ssection: autocorrelation function}

\begin{figure}
\begin{center}
\includegraphics[width=8.6cm]{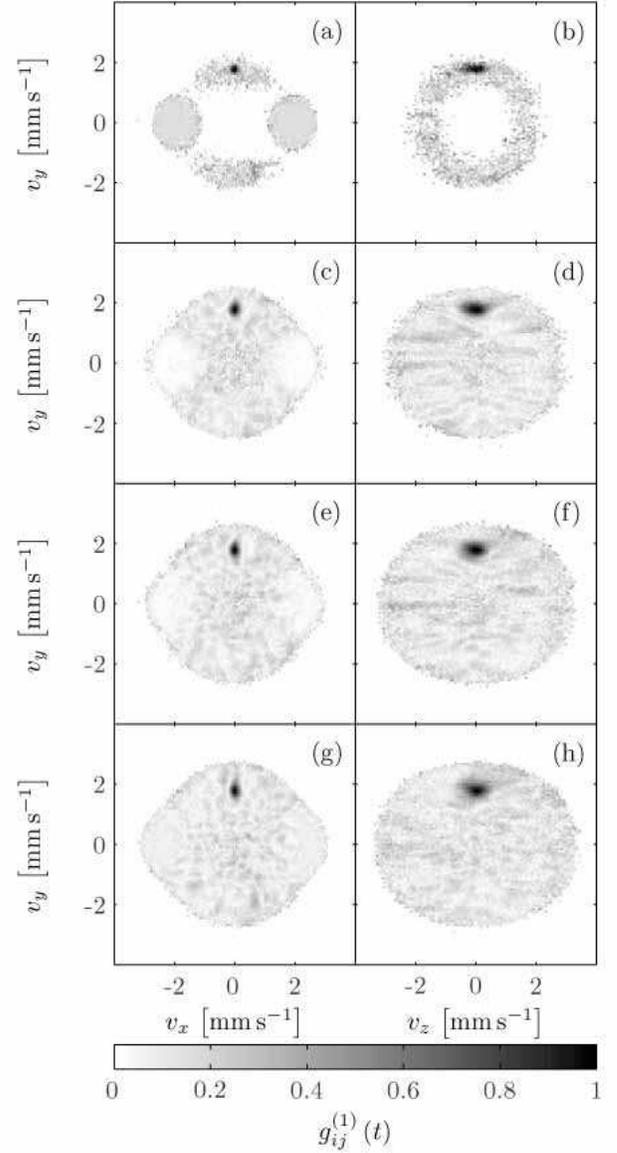}
\end{center}
\vspace{-0.5cm}
\caption{Normalized mode space autocorrelation function $g^{\left( 1
\right)}_{ij} \left( t \right)$ on the planes $v_z = 0$ (a,c,e,g) and $v_x = 0$
(b,d,f,h) at 6.6~ms (a,b), 13.1~ms (c,d), 19.7~ms (e,f) and 37.7~ms (g,h). Here
${\bf v}_i = ( 0, 1.8~{\rm mm\,s}^{-1}, 0)$ and results are only
shown for modes with $N_j > 1/4$.}
\label{Figure: g11 autocorrelation}
\end{figure}

As shown by Figs.~\ref{Figure: single run k slices}, \ref{Figure: single run
higher k} and \ref{Figure: single run phase only}, the scattering halo in
momentum space is characterized by discrete phase grains. To quantify the size
of these grains, and hence the range of phase order within the halo, we use the
normalized autocorrelation function
\begin{equation}
g^{\left( 1 \right)}_{ij} \left( t \right) \equiv \frac{\left \langle
\hat{a}_i^{\dagger} \hat{a}_j \right \rangle \left( t \right)}{\sqrt{\left
\langle \hat{a}_i^{\dagger} \hat{a}_i \right \rangle \left( t \right) \left
\langle \hat{a}_j^{\dagger} \hat{a}_j \right \rangle \left( t \right)}},
\end{equation}
which in terms of moments of the Wigner function is
\begin{equation}
g^{\left( 1 \right)}_{ij} \left( t \right) = \frac{\left \langle \alpha_i^*
\left( t \right) \alpha_j \left( t \right) \right \rangle_W - \frac{1}{2}
\delta_{i,j}}{\sqrt{\left[ \left \langle \left| \alpha_i \left( t \right)
\right|^2 \right \rangle_W - \frac{1}{2} \right] \left[ \left \langle \left|
\alpha_j \left( t \right) \right|^2 \right \rangle_W - \frac{1}{2} \right]}}.
\end{equation}
Although modes $i$ and $j$ can be arbitrarily chosen, for this analysis we fix
mode $i$ at ${\bf v}_i = \left( 0, 1.8~{\rm mm\,s}^{-1}, 1 \right)$ and vary $j$
over the full mode space, for which the autocorrelation function is shown on
the planes $v_z = 0$ and $v_x = 0$ in Fig.~\ref{Figure: g11 autocorrelation}.
From these plots we observe that the significant feature of the autocorrelation
function is a small ellipsoid, visible as a black grain centered on ${\bf v}_j =
{\bf v}_i$, whose size and shape remains relatively unchanged through time.

To more precisely quantify the time-dependent shape of this grain we have
fitted a three-dimensional Gaussian profile to the correlation function, over a
region of approximately the same extent as the grain. At the time when
significant population is first established in the halo (8~ms), the HWFM widths
of the fitted Gaussian are approximately 0.3~mm\,s$^{-1}$, 0.3~mm\,s$^{-1}$ and
0.8~mm\,s$^{-1}$ in the $v_x$, $v_y$ and $v_z$ directions respectively. These
widths reflect the asymmetry of the initial condensate wavepackets (in velocity
space), and are approximately 1.5 times as large as the velocity radii of the
condensate wavepackets in the Thomas-Fermi approximation (being 0.19~mm\,s$^{-1}$
in the $v_x$ and $v_y$ directions and 0.54~mm\,s$^{-1}$ in the $v_z$ direction).
At later times, the fitted widths are found to increase anisotropically, 
so
that by 40~ms the fitted widths are approximately 0.3~mm\,s$^{-1}$,
0.6~mm\,s$^{-1}$ and 0.9~mm\,s$^{-1}$. These increases are driven by scattering
events, and reflect the inverse extent of the colliding system in coordinate
space (see Fig.~\ref{Figure: single run x slices}).

\subsection{Parameter dependence}

In this section we explore the dependence of the coherent and incoherent
populations on the initial condensate number and on the initial relative
wavepacket speed. These parameters, which are relatively easily adjusted
experimentally, are the only ones we change --- all other simulation
parameters, including $U_0$, $\omega_{x,y,z}$ and the details of the simulation
grids, are held fixed. We obtain the time-dependent total coherent and
incoherent populations for each parameter set using the extrapolation method
outlined in section~\ref{sssection: total populations}. For all parameter sets
two trajectories were simulated.

\begin{figure}
\begin{center}
\resizebox{8.5cm}{!}{\includegraphics[width=8.6cm]{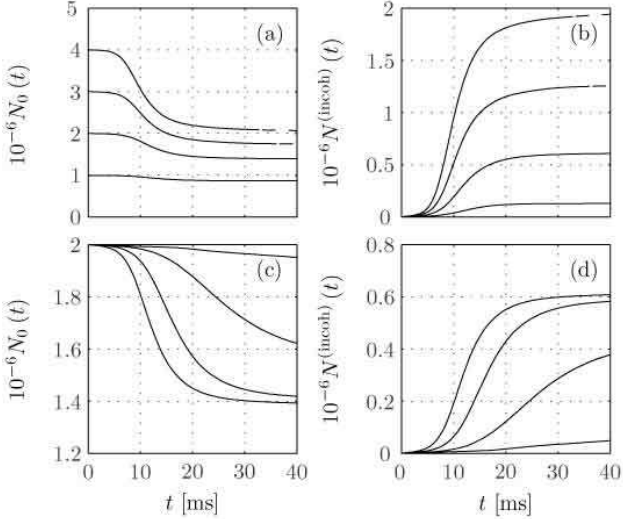}}
\end{center}
\vspace{-0.5cm}
\caption{Total coherent and incoherent populations calculated using the
extrapolation method for varying collision parameters. Plots (a,b) show
(respectively) the coherent and incoherent populations for collisions with
$\Delta v = 4.0$~mm\,s$^{-1}$ with (lowest to highest) $N_0 \left( t = 0 \right)
= \left\{ 1,2,3,4 \right\} \times 10^6$. Plots (c,d) show the coherent and
incoherent populations for collisions with $N_0 \left( t = 0 \right) = 2 \times
10^6$ and (highest to lowest in (c) and lowest to highest in (d)) $\Delta v =
\left\{ 1.0, 2.0, 3.0, 4.0 \right\}$~mm\,s$^{-1}$.}
\label{Figure: populations varying parameters}
\end{figure}

\subsection{Initial condensate number}

In Fig.~\ref{Figure: populations varying parameters}~(a,b) we plot the total
coherent and incoherent populations for collisions at $\Delta v =
4.0$~mm\,s$^{-1}$ with total initial condensate numbers of $N_0 \left( t = 0
\right) = \left\{ 1,2,3,4 \right\} \times 10^6$. From these curves we can see
that by increasing the total number of particles within the system, the amount
of incoherent population generated by the collision also increases. We find that
the peak rate of incoherent particle formation increases roughly linearly with
$N_0 \left( 0 \right)$, and that the time that this peak occurs decreases, with
increasing particle number. The dashed parts of the curves for $N_0 \left( t =
0 \right) = \left\{ 3,4 \right\} \times 10^6$ indicate results where the real
particle density has begun to exit the simulation volume.

\subsection{Collision speed}

In Fig.~\ref{Figure: populations varying parameters}~(c,d) we plot the total
coherent and incoherent populations for collisions with $N_0 \left( t = 0
\right) = 2 \times 10^6$ and $\Delta v = \left\{ 1.0, 2.0, 3.0, 4.0
\right\}$~mm\,s$^{-1}$. These curves show that the rate of incoherent population
scattering increases (again found to be roughly linearly) with increasing
relative wavepacket speed, a result which is expected from classical
collisional theory $\left( n \sigma v \right)$. However, as can be seen for the
curves with $\Delta v = \left\{ 3.0,4.0 \right\}$~mm\,s$^{-1}$, the total amount
of incoherent population generated is less strongly dependent upon $\Delta v$,
as the overlap time of the condensate wavepackets, and hence the period over
which the major mechanism of incoherent particle formation acts, scales as
$\approx 1/\sqrt{\Delta v}$. For the lower values of $\Delta v$, broadening of
the condensate wavepackets leads to the population exiting the simulation
volume before packet separation. Note that the the speed of sound for this
system (taken as for a homogeneous system at the peak density of the condensate
at $t = 0$) is 1.5~mm\,s$^{-1}$.

\subsection{The halo formation process}

The basic mechanism of halo formation is pairwise scattering of condensate
particles into two distinct halo modes, which can be viewed as a four-wave
mixing process. However, not all modes within the halo region exhibit
population growth, a feature that can be understood in terms of the phase
relationships required for growth. For plane wave modes, the nonlinear portion
of the mode amplitude evolution is
\begin{equation}
\label{nonlinear SDE portion 1}
i \frac{d\alpha_j}{dt}^{\left( {\rm nonlin} \right)} = \frac{U_0}{V}
\sum_{rst \in L} \alpha_r^* \alpha_s \alpha_t \delta_{{\bf k}_j + {\bf k}_r,
{\bf k}_s + {\bf k}_t},
\end{equation}
representing a process where $s + t \rightarrow j + r$. Writing the mode
amplitudes as $\alpha_j \left( t \right) = \sqrt{\mathcal{N}_j} \exp \left[  i
\theta_j \left( t\right) \right]$, the rate of population change for a single
mode (in the absence of external potentials) is
\begin{equation}
\label{population change from SDE}
\frac{d\mathcal{N}_{j}}{dt}=-\frac{2U_{0}}{V}\sum_{rst\in L}\sqrt
{\mathcal{N}_{j}\mathcal{N}_{r}\mathcal{N}_{s}\mathcal{N}_{t}}\sin\left(
\Theta_{jrst}\right)  \delta_{\mathbf{k}_{j}+\mathbf{k}_{r},\mathbf{k}
_{s}+\mathbf{k}_{t}},
\end{equation}
where $\Theta_{jrst}\equiv\theta_{j}+\theta_{r}-\theta_{s}-\theta_{t}$. The
phase evolution corresponding to Eq.~(\ref{population change from SDE}) is
\begin{equation}
\label{phase change from SDE}
\frac{d\theta_{j}}{dt}=-\frac{U_{0}}{V}\sum_{rst\in L}\sqrt{\frac
{\mathcal{N}_{r}\mathcal{N}_{s}\mathcal{N}_{t}}{\mathcal{N}_{j}}}\cos\left(
\Theta_{jrst}\right)  \delta_{\mathbf{k}_{j}+\mathbf{k}_{r},\mathbf{k}
_{s}+\mathbf{k}_{t}}.
\end{equation}
At early times, the evolution of the halo modes is driven almost entirely by
the condensate wavepackets, with the dominant processes being scattering events
involving the destruction of one quanta from each wavepacket. The role of the
phase $\Theta_{jrst}$ is critical, as can be understood from a simplified model
of four modes only, two highly occupied modes $s$ and $t$ representing the two
condensate packets, and a pair of scattered modes $j$ and $r$ of lower
occupation, for which $\mathbf{k}_{r}=-\mathbf{k}_{j}$ (in the centre of mass
frame). Initially the phase $\Theta_{jrst}$ is randomly set, and if
$-\pi<\Theta_{jrst}<0$, population transfers from the pair $(s,t)$ into the
pair $(j,r),$ while if $0<\Theta_{jrst}<\pi,$ the population transfer is in the
opposite direction. The phase $\Theta_{jrst}$ evolves towards either $-\pi/2$
(maximum gain for scattered modes) or $+\pi/2$ (maximum loss), where it
stabilizes while population is still available for transfer. Thus some pairs
$(j,r)$ will grow, while others will decrease in size. If
$\mathcal{N}_{j}\ll\mathcal{N}_{r}$, the phase $\theta_{j}$ can change rapidly
(see Eq.~(\ref{phase change from SDE})), possibly leading to a change in the
direction of population transfer.

For the multimode case, the main additional feature is that the scattered modes
no longer need be precise momentum opposites, because of the range of momentum
modes available in the condensate packets. Thus for a particular scattered mode
$j$ there is a range of possible modes near its conjugate momentum mode $r$
which can be convolved with the condensate packets to contribute to the growth
(or loss) in $j$, (see Eq.~(\ref{nonlinear SDE portion 1})). Labelling this set
of modes about $r$ as $R$, then from the properties of convolution, the size of
$R$ is about twice that of a condensate packet. There is of course a
corresponding set of modes $J$ about $j$, and if the phases and populations for
a pair of modes from $R$ and $J$ are favorable for growth, these phases can
quickly be locked across the regions $R$ and $J$ (see Fig ~\ref{Figure: single
run phase only}). Population growth is a stimulated process, so while $R$ and
$J$ will now grow rapidly, regions which have not established a favorable phase
are left behind. 

We note that this discussion helps explain the incomplete pair correlation
observed for halo modes $j$ and $r$, because the scattered pair are not
required to have exactly opposite momentum, due to the range of momentum modes
available in the condensate packets.

\section{Conclusion}

In this paper we have presented a complete derivation of the truncated Wigner
method, paying particular attention to the limits of validity of the Wigner
truncation. Using this formalism, we have presented simulation results, both
from a single trajectory and from ensembles of trajectories, of collisions
between distinct Bose-Einstein condensate wavepackets. This formalism includes
both stimulated \emph{and} spontaneous processes, allowing for a complete
treatment of system dynamics. In particular we have observed the generation of
highly populated $s$-wave scattering haloes. Previous treatments of similar
collisional systems have been restricted to the low-scattering limit, and have
not provided a complete description of the field.

The most significant process not included in our treatment of these collisions
is three-body recombination \cite{Wieman1997a}. In a further paper
\cite{Norrie2005c} we extend the truncated Wigner method to include this
process, and investigate its effect on a simple system. For the colliding
systems considered in this paper we have found using this extended formalism
that three-body recombination has a negligible effect on the dynamics.

\subsection{Relationship to the projected \GPE}

The various forms of the truncated Wigner differential equation,
Eqs.~(\ref{Mode space SDE},\ref{Hybrid SDE},\ref{Coordinate space SDE}), are
functionally identical to the Projected \GPE\ of Davis \emph{et al.}
\cite{Davis2001a,Davis2001b,Davis2002a}. However the projected  \GPE\ treatment
has been used only for relatively high temperature situations, in which the
thermal fluctuations are much larger than the quantum fluctuations, whereas in
the truncated Wigner method the quantum mechanical nature of the system is
still largely present in the form of mode amplitude fluctuations in the initial
state (see below). The initial state of the \GPE\ system would represent the
limit, in the truncated Wigner function method, in which the size of the
quantum fluctuations becomes negligible in comparison with the size of the
order parameter of the field, while keeping the number of particles in the
field constant. Thus the \GPE\ can be considered to represent the thermodynamic
limit of the truncated Wigner approach where spontaneous, and spontaneously
initiated, processes are far less important than stimulated processes. Of
course the alternate approach is then simply to take the \GPE, and ``seed'' each
mode with a certain amount of noise, such that previously unavailable
spontaneous processes become possible. While this approach would avoid some of
the more complicated aspects of the truncated Wigner method, one could not then
make the unambiguous connections to the underlying quantum nature of the system
which the truncated Wigner method allows.



\begin{thebibliography}{10}
    
\bibitem{Chikkatur2000a}
A.~P. Chikkatur {\it et~al.}, Phys. Rev. Lett. {\bf 85},  483  (2000).

\bibitem{Katz2002a}
N. Katz, J. Steinhauer, R. Ozeri, and N. Davidson, Phys. Rev. Lett. {\bf 89},
  220401  (2002).

\bibitem{Vogels2002a}
J.~M. Vogels, K. Xu, and W. Ketterle, Phys. Rev. Lett. {\bf 89},  020401
  (2002).

\bibitem{Band2000a}
Y.~B. Band, M. Trippenbach, J. J.~P.~Burke, and P.~S. Julienne, Phys. Rev.
  Lett. {\bf 84},  5462  (2000).


\bibitem{Bach2002a}
R. Bach, M. Trippenbach, and K. Rz\c{a}\.{z}ewski, Phys. Rev. A {\bf 65},
  063605  (2002).

\bibitem{Yurovsky2002a}
V.~A. Yurovsky, Phys. Rev. A. {\bf 65},  033605  (2002).




\bibitem{Norrie2005a}
A.~A. Norrie, R.~J. Ballagh, and C.~W. Gardiner, Phys. Rev. Lett. {\bf 94},
  040401  (2005).

  
\bibitem{Braaten1999a}
E. Braaten and A. Nieto, Eur. Phys. J. {\bf B11},  143  (1999)
  [cond-mat/9707199].

  
\bibitem{Andersen2004a}
J.~O. Andersen, Rev. Mod. Phys. {\bf 76},  599  (2004).

\bibitem{Steel1998a}
M.~J. Steel {\it et~al.}, Phys. Rev. A {\bf 58},  4824  (1998).

\bibitem{Deuar2006a}
P. Deuar and P. D. Drummond, J. Phys.  A: Math.  Gen.
{\bf  39} 1163 (2006)

\bibitem{Deuar2005a}
P. Deuar and P. D. Drummond, cond-mat/0501058 (2005)


\bibitem{Stoof1999a}
H.~T.~C. Stoof, J. Low Temp. Phys. {\bf 114},  11  (1999).

\bibitem{Duine2001a}
R.~A. Duine and H.~T.~C. Stoof, Phys. Rev. A {\bf 65},  013603  (2001).

\bibitem{Stoof2001a}
H.~T.~C. Stoof and M.~J. Bijlsma, J. Low Temp. Physics {\bf 124},  431  (2001).

\bibitem{Davis2001b}
M.~J. Davis, R.~J. Ballagh, and K. Burnett, J. Phys. B {\bf 34},  4487  (2001).

\bibitem{Davis2001a}
M.~J. Davis, S.~A. Morgan, and K. Burnett, Phys. Rev. Lett. {\bf 87},  160402
  (2001).

\bibitem{Sinatra2001a}
A. Sinatra, C. Lobo, and Y. Castin, Phys. Rev. Lett. {\bf 87},  210404  (2001).

\bibitem{Goral2001a}
K. G\`{o}ral, M. Gajda, and K. Rz\c{a}\.{z}ewski, Opt. Express {\bf 8},  92
  (2001).

\bibitem{Gardiner2002a}
C.~W. Gardiner, J.~R. Anglin, and T.~I.~A. Fudge, J. Phys. B {\bf 35},  1555
  (2002).

\bibitem{Davis2002a}
M.~J. Davis, S.~A. Morgan, and K. Burnett, Phys. Rev. A {\bf 66},  053618
  (2002).

\bibitem{Goral2002a}
K. G\'{o}ral, M. Gajda, and K. Rz\c{a}\.{z}ewski, Phys. Rev. A {\bf 66},
  051602(R)  (2002).

\bibitem{Schmidt2003a}
H. Schmidt {\it et~al.}, J. Opt. B {\bf 5},  96  (2003).

\bibitem{Katz2005a}
N. Katz, E. Rowen, R. Ozeri and N. Davidson, Phys. Rev. Lett. 
{\bf  95}, 220403 (2005)

\bibitem{Drummond1999a}
P.~D. Drummond and J.~F. Corney, Phys. Rev. A {\bf 60},  R2661  (1999).

\bibitem{Gardiner2000a}
C.~W. Gardiner and P. Zoller, {\em Quantum Noise}, 3rd ed. (Springer-Verlag,
  Berlin, 2004).

\bibitem{Morgan2000a}
S.~A. Morgan, J. Phys. B {\bf 33},  3847  (2000).

\bibitem{Polkovnikov2003a}
A. Polkovnikov, Phys. Rev. A {\bf 68},  053604  (2003).

\bibitem{Sinatra2002a}
A. Sinatra, C. Lobo, and Y. Castin, J. Phys. B {\bf 35},  3599  (2002).

\bibitem{Gardiner2003b}
C.~W. Gardiner, {\em Handbook of Stochastic Methods}, 3rd ed. (Springer-Verlag,
  Berlin, 2004).

\bibitem{Ballagh2000a}
R.~J. Ballagh, Computational methods for nonlinear partial differential
  equations, \url{ http://www.physics.otago.ac.nz/research/uca/resources/
  comp_lectures_ballagh.html}, 2000.

\bibitem{Press1992a}
W.~H. Press, S.~A. Teukolsky, W.~T. Vetterling, and B.~P. Flannery, {\em
  Numerical Recipes in C}, 2nd ed. (Cambridge University Press, Cambridge, MA,
  1992).

\bibitem{Ketterle2002a}
J.~M. Vogels, K. Xu, and W. Ketterle, Phys. Rev. Lett. {\bf 89},  020401
  (2002).

\bibitem{Phillips1999a}
M. Kozuma {\it et~al.}, Phys. Rev. Lett. {\bf 82},  871  (1999).

\bibitem{Stringari1999a}
F. Dalfovo, S. Giorgini, L.~P. Pitaevskii, and S. Stringari, Rev. Mod. Phys.
  {\bf 71},  463  (1999).

\bibitem{Morgan1999a}
S.~A. Morgan, A Gapless Theory of Bose-Einstein Condensation in Dilute Gases at
  Finite Temperatures, DPhil Thesis, University of Oxford, 1999.
\bibitem{Wieman1997a}
E.~A. Burt {\it et~al.}, Phys. Rev. Lett. {\bf 79},  337  (1997).

\bibitem{Norrie2005c}
A.~A. Norrie, R.~J. Ballagh, C.~W. Gardiner, and A.~S. Bradley, Submitted.

\end{thebibliography}
\end{document}